\documentclass[12pt]{article}

\usepackage{verbatim,color,amssymb,amsmath}
\usepackage{mathtools}
\usepackage{amsthm}		
\usepackage{bm}
\usepackage{dsfont}			
\usepackage{multirow}
\usepackage[mathscr]{euscript}
\usepackage{fancyhdr}
\usepackage{enumitem}
\usepackage{graphicx}
\usepackage{geometry}
\usepackage{xcolor}
\usepackage{hyperref}
\usepackage{bbm}
\usepackage[section]{placeins}
\usepackage{cleveref}

\usepackage[T1]{fontenc}   

\usepackage[ruled, norelsize, vlined]{algorithm2e}
\SetKwInput{kwInit}{Initialization}
\SetKwInput{kwOpt}{Optimization}
\SetKwInput{kwComp}{Final computation of quantities of interest}

\usepackage{booktabs}
\usepackage{textcomp}
\usepackage{manyfoot}

\usepackage{subfig}
\usepackage{float}
\usepackage{lineno}

\usepackage{setspace}
\usepackage{tikz}
\usetikzlibrary{arrows}
\usepackage{wrapfig}

\usepackage[style=ext-authoryear, maxcitenames=1, giveninits=true, maxbibnames=99, sortcites=false, natbib]{biblatex} 

\renewbibmacro{in:}{%
  \ifentrytype{article}
    {}
    {\bibstring{in}%
     \printunit{\intitlepunct}}}
\DeclareFieldFormat[article]{pages}{#1}

\usepackage{lscape}

\newtheorem*{Proof*}{Proof}

\newtheorem{Prop}{\underline{\bf Proposition}}
\newtheorem{Lem}{\underline{\bf Lemma}}

\newcommand{\R}{{\mathbb{R}} }
\newcommand{\E}{{\mathbb{E}} }
\newcommand{\Var}{{\operatorname{var} }}
\newcommand{\new}{{\textsc{new}} }

\newcommand{\diag}{{\text{diag}} }
\newcommand{\ep}{{\textsc{ep}} }
\newcommand{\bzero}{ {\bf 0} }

\newcommand{\bmm}{ {\bf m} }
\newcommand{\br}{ {\bf r} }
\newcommand{\bt}{ {\bf t} }

\newcommand{\bx}{ {\bf x} }

\newcommand{\bS}{ {\bf S} }

\newcommand{\by}{ {\bf y} }
\newcommand{\bX}{ {\bf X} }

\newcommand{\bv}{ {\bf v} }

\newcommand{\bI}{ {\bf I} }
\newcommand{\bQ}{ {\bf Q} }
\newcommand{\bV}{ {\bf V} }

\newcommand{\bw}{ {\bf w} }

\newcommand{\bC}{ {\bf C} }
\newcommand{\bK}{ {\bf K} }

\newcommand{\bLambda}{ {\boldsymbol \Lambda} }

\newcommand{\bbeta}{ {\boldsymbol \beta} }
\newcommand{\bmu}{ {\boldsymbol \mu} }

\newcommand{\bSigma}{ {\boldsymbol \Sigma} }

\newcommand{\bOmega}{ {\boldsymbol \Omega} }
\newcommand{\bomega}{ {\boldsymbol \omega} }
\newcommand{\balpha}{ {\boldsymbol \alpha} }
\newcommand{\bxi}{ {\boldsymbol \xi} }

\newcommand{\btheta}{ {\boldsymbol \theta} }

\usepackage{comment}
\newcommand{\cmmnt}[1]{\ignorespaces}

\allowdisplaybreaks

\begin{document}
\newrefsection
	\thispagestyle{empty}
	\baselineskip=28pt

	\begin{center}
		{\LARGE{\bf Scalable expectation propagation\\ for generalized linear models
		}}
	\end{center}
\baselineskip=12pt
\vskip 2mm

\begin{center}
    Niccolò Anceschi$^{1}$ (niccolo.anceschi@duke.edu)\\
    Augusto Fasano$^{2}$ (augusto.fasano@unicatt.it)\\
    Beatrice Franzolini$^{3}$ (beatrice.franzolini@unibocconi.it)\\
    Giovanni Rebaudo$^{4}$ (giovanni.rebaudo@unito.it)
    \vskip 3mm
    $^{1}$Duke University, Durham, USA\\
    $^{2}$Catholic University of the Sacred Heart, Milan, Italy\\
    $^{3}$Bocconi University, Milan, Italy\\
    $^{4}$University of Torino, Turin, Italy
\end{center}
	
\vskip 8mm
\begin{center}
	{\Large{\bf Abstract}} 
\end{center}
Generalized linear models (\textsc{glm}s) arguably represent the standard approach for statistical regression beyond the Gaussian likelihood scenario.
When Bayesian formulations are employed, the general absence of a tractable posterior distribution has motivated the development of deterministic approximations, which are generally more scalable than sampling techniques.
Among them, expectation propagation (\textsc{ep}) showed extreme accuracy, usually higher than many variational Bayes solutions.
However, the higher computational cost of \ep posed concerns about its practical feasibility, especially in high-dimensional settings.
We address these concerns by deriving a novel efficient formulation of \ep for \textsc{glm}s, whose cost scales linearly in the number of covariates $p$. This reduces the state-of-the-art $\mathcal{O}(p^{2} n)$ per-iteration computational cost of the \ep routine for \textsc{glm}s to $\mathcal{O}(p n \cdot \min\{p,n\})$, with $n$ being the sample size.
We also show that, for binary models and log-linear \textsc{glm}s approximate predictive means can be obtained at no additional cost.
To preserve efficient moment matching for count data, we propose employing a combination of log-normal Laplace transform approximations, avoiding numerical integration.
These novel results open the possibility of employing \ep in settings that were believed to be practically impossible.
Improvements over state-of-the-art approaches are illustrated both for simulated and real data.
The efficient \textsc{ep} implementation is available at \href{https://github.com/niccoloanceschi/EPglm/}{https://github.com/niccoloanceschi/EPglm}.
	
\baselineskip=12pt
\vskip 8mm

\noindent\underline{\bf Key Words}: 
Generalized Linear Model,
Expectation Propagation,
Bayesian Inference,
Extended Multivariate Skew-Normal Distribution,
Unified Skew-Normal Distribution.
	
\par\medskip\noindent
\underline{\bf Short/Running Title}: 
Scalable EP for GLMs.


\clearpage\pagebreak\newpage
\pagenumbering{arabic}
\newlength{\gnat}
\setlength{\gnat}{16pt}  
\baselineskip=\gnat

\section{Introduction}\label{sec: intro}

Generalized linear models (\textsc{glm}s) arguably represent the most popular off-the-shelf tool for data regression for a wide variety of data, with binary and count responses representing their main application areas.
The lack of a tractable posterior distribution in their Bayesian formulation has motivated notable efforts to develop computational methods that can efficiently scale to high-dimensional settings (large $p$) or large datasets (large $n$), with the former providing a primary area of interest.
See, for instance, the review paper by \citet{chopin2017leave} about the big $p$ problem faced by the available computational methods for Bayesian binary regression computations.
Among the scalable deterministic approximate methods, a central role has certainly been played by variational Bayes (\textsc{vb}) approximations \citep[e.g.,][]{blei2017variational}, which are usually much faster than sampling-based techniques.
This comes however at the cost of possible accuracy issues in certain settings, making some of the \textsc{vb} approximations a sound replacement for the exact posterior only in some scenarios.
This is sub-par compared to having an approximation scheme offering reliable performances regardless of the specific setting considered.
Empirically, expectation propagation (\textsc{ep}) \citep{minka2001expectation, minka2001family} has demonstrated an improved quality in posterior approximation for a wide variety of models, including binary \citep{chopin2017leave}, tobit \citep{anceschi2023bayesian}, Poisson \citep{zhang2019expectation}, heteroscedastic, lasso-penalized, and quantile regression \citep{zhou2023fast}, as well as for hierarchical models more in general \citep{vehtari2020expectation}.
Indeed, when testing its performance with logit likelihood and various priors on datasets of moderate size, the empirical approximation error turned out to be so small that \citet{chopin2017leave} conclude that \textit{it seems that \textsc{ep} may be safely used as a complete replacement of sampling-based methods} (Section 5.1.1).
However, such improved accuracy comes at a higher computational cost than \textsc{vb} methods, thus motivating questions about the applicability of this excellent approximate method in large $p$ settings, namely $p\gg 1000$ \citep{chopin2017leave}.
To address this issue we develop a novel \textsc{ep} formulation for Bayesian \textsc{glm}s with multivariate Gaussian prior which scales linearly in the number of covariates $p$.
This improves over current state-of-the-art implementations by multiple orders of magnitude and makes \textsc{ep} a feasible option in \textsc{glm}s with dimensions well beyond the above threshold, paving the way to the employment of \textsc{ep} in settings where available sampling methods are impractical and other available approximations are less accurate.

In \textsc{glm}s, the effect of the $p$-dimensional vector of predictors $\bx=(x_{1},\ldots,x_{p})^{\intercal}$ on the distribution of the response variable $y$ is modelled through a \textit{link function} $g(\cdot)$, which relates the linear predictor $\eta=\bx^{\intercal} \bbeta$ to the mean of the response variable via the relation $\E[y\mid \bbeta]=g^{-1}(\eta)$, with $\bbeta$ a $p$-dimensional unknown parameter.
For binary regression models with Bernoulli likelihood, classical link functions are represented by the logit or probit links, while for count data default specifications combine a Poisson or Negative Binomial likelihood with $\log(\cdot)$ link function.
Other models have also been studied, including, for instance, Gamma regression for non-negative continuous responses, where a Gamma likelihood is usually combined with a $\log(\cdot)$ link function.
When considering their Bayesian formulation, the model is completed by specifying a prior distribution for the parameter $\bbeta$, with the zero-mean spherical multivariate Gaussian distribution constituting a default choice.
Unfortunately, a tractable closed-form posterior distribution for the \textsc{glm}s is generally lacking beyond the Gaussian regression.

State-of-the-art Bayesian approaches usually rely on posterior sampling methods --- see, e.g., \citet{holmes2006bayesian,polson2013bayesian,durante2019conjugate,zhang2021large,zens2023ultimate} and the seminal contribution by \citet{albert1993bayesian} for binary regression models, or \citet{fruhwirth2009improved,chanialidis2018efficient, dangelo2023efficient} for the Poisson-like models --- or more scalable approximate posterior inference --- see, e.g., \citet{jaakkola2000bayesian,consonni2007meanfield,durante2019conditionally,fasano2022scalable} for binary regression models, or \citet{arridge2018variational,zhang2019expectation} for Poisson regression.
Most of the above-mentioned sampling methods rely on \textsc{mcmc} schemes, often introducing a data augmentation step, in the spirit of the well-known Gibbs sampler for the probit model originally developed by \citet{albert1993bayesian}, who exploited the augmented variables to obtain conjugate full conditionals.
These approaches make in principle posterior inference feasible for a wide variety of models, but the possible high correlation between the augmented data and the parameters can cause convergence issues and thus unreliable results, especially in unbalanced datasets \citep{johndrow2018mcmc}.
Other sampling alternatives, like the no-U-turn Hamiltonian sampler by \citet{hoffman2014nuts} or adaptive Metropolis-Hastings techniques \citep{haario2001adaptive,roberts2009examples}, are possible.
Although these approaches might mitigate autocorrelation among samples and improve mixing, they remain computationally impractical for high-dimensional problems \citep{durante2019conjugate, fasano2022scalable}.
A notable exception is represented by \citet{durante2019conjugate}, who developed an i.i.d.\ sampler for the posterior distribution in Bayesian probit models.
This leverages on the representation of the $\textsc{sun}_{p,n}$ posterior \citep{arellano2006unification} as a linear combination of a $p$-variate Gaussian and an $n$-variate truncated Gaussian, making it particularly well suited for high-dimensional problems.
However, the need to sample from truncated multivariate Gaussians whose dimension equals the number of the observations represents the computational challenge of such an algorithm: despite recent promising results on the topic \citep{pakman2014exact,botev2017normal, genton2018hierarchical, zhang2021large}, i.i.d.\ posterior samples are hard to get when $n$ is above a few hundred.

More scalable deterministic approximations were thus developed to circumvent the computational issues mentioned above.
However, in high-dimensional settings, some of the most popular approximate methods, like mean-field \textsc{vb} or Laplace approximation, may turn out to be unreliable in high dimensions \citep{fasano2022scalable, fasano2022class, anceschi2023bayesian}.
In binary and multinomial Bayesian regression, the above contributions overcame such drawbacks by developing a partially-factorized variational (\textsc{pfm-vb}) approximation of the posterior distribution which, for any fixed $n$, converges to the true posterior distribution as $p$ diverges under reasonable assumptions.
Moreover, \textsc{pfm-vb} is always guaranteed to improve over mean-field approximations \citep{consonni2007meanfield, girolami2006variational}.
The computational costs are orders of magnitude smaller than the ones of sampling methods, and, empirically, the approximate posterior moments closely match the ones obtained via i.i.d.\ sampling for $p/n\ge 2$.
Other variational approximations have been developed for the logit \citep{jaakkola2000bayesian} and Poisson \citep{arridge2018variational} to get fast posterior approximations.
See also \cite{chopin2017leave} for a complete review of computational methods for Bayesian binary regression, covering also  Laplace and improved nested Laplace approximation (\textsc{inla}) \citep{rue2009approximate}, as well as other exact methods whose review is beyond the scope of the paper.
The overall accuracy of the \textsc{vb} approaches is still an open area of research and in many cases, it is very setting-specific.
In general, \textsc{ep} provides better posterior approximations across scenarios for \textsc{glm}s \citep{chopin2017leave, anceschi2023bayesian, zhang2019expectation}.
This comes, however, at a higher computational cost, which makes current implementations not feasible for settings where $p>1000$.
We show that an efficient derivation of \textsc{ep} is possible, having linear cost in $p$ and reducing the $\mathcal{O}(p^{2} n)$ per-iteration cost reported in \citet{chopin2017leave} to $\mathcal{O}(pn\cdot \min\{p,n\})$.
This improvement makes \textsc{ep} feasible in settings that are out of reach for state-of-the-art implementations.
The resulting efficient implementation can be fully leveraged to provide accurate approximations of the marginal likelihood, which is often overlooked in the literature.
Besides such structural enhancement, we achieve further computational improvement for count data, by proposing a novel approximate moment matching step for Poisson regression.
Similarly to standard practice in the logit case, this avoids the need for numerical integration, benefiting both in terms of efficiency and stability.

The rest of the article is organized as follows.
Section \ref{sec: EP} reviews the \ep algorithm and its employment to compute the approximate posterior moments and marginal likelihood.
Section \ref{sec: EPglm} details our efficient \ep routines for various \textsc{glm}s, while Section \ref{sec: EPhybrid} shows multiple approximations of the hybrid moments required by the \ep algorithm, avoiding numerical quadrature.
We also show that in several relevant cases, the required hybrid moments and/or the predictive means are available in closed form.
Section \ref{sec: simstudies} reports the results of our proposed algorithm over simulated data, while Section \ref{sec: illustration} presents the performance of the proposed method over real data applications.
Section \ref{sec: discussion} contains concluding remarks and further interesting research questions arising from our findings.
The code to reproduce the results is available at \href{https://github.com/niccoloanceschi/EPglm/}{https://github.com/niccoloanceschi/EPglm}.

\section{Expectation propagation}\label{sec: EP}
In this section, we briefly review the principles and the main steps of expectation propagation (\textsc{ep}) \citep{minka2001expectation,minka2001family} for Bayesian inference in a generic regression model
\begin{equation}
\label{eq: BayesianModel}
    y_{i} \mid \bbeta \overset{ind}{\sim} p(y_{i}\mid\bbeta)\quad i=1,\ldots,n,\quad\bbeta\sim p(\bbeta),
\end{equation}
where the likelihood contribution $p(y_{i}\mid\bbeta)$ depends on the unknown parameter $\bbeta$ and a known covariate vector $\bx_{i}\in\R^p$, which is thus removed from the conditioning part.
Throughout the paper, we will assume univariate $y_{i}$'s and the well-established zero-mean spherical multivariate Gaussian prior for the parameter $\bbeta$, although computations trivially adapt to more general cases.
We will then take $p(\bbeta)=\phi_{p}(\bbeta;\nu^{2} \bI_{p})$, where $\bI_{p}$ is the identity matrix of dimension $p$ and $\phi_{p}(\bt;\bS)$ denotes the density of a $p$-variate Gaussian random variable with mean $\boldsymbol{0}_p$ and covariance matrix $\bS$, evaluated at $\bt$.

\subsection{Approximation of posterior moments}
Calling $\by=(y_{1},\ldots,y_{n})$, the \textsc{ep} framework approximates the posterior distribution $$p(\bbeta\mid\by)\propto p(\bbeta)\prod_{i=1}^{n} p(y_{i}\mid \bbeta)$$ with a density $q_{\ep}(\bbeta)$ having the same factorization form $$ q_{\ep}(\bbeta) \propto q_{0}(\bbeta) \prod_{i=1}^{n}q_{i}(\bbeta),$$
where the global approximation $q_{\ep}(\bbeta)$ and the factors (also referred to as \textit{sites}) $q_{i}(\bbeta)$, $i=0,\ldots,n$, are taken from a possibly unnormalized exponential family, as clarified below.
The key operating principle of the \textsc{ep} scheme is to refine iteratively each approximate site $q_{i}(\bbeta)$ against the exact one $p(y_{i}\mid \bbeta)$, exploiting the information coming from the remaining factors $\{q_{j}(\bbeta)\}_{j \neq i}$.
This helps in targeting regions of high posterior density and, thus, constructing a globally accurate approximation $q_{\ep}(\bbeta)$.

When the parameter $\bbeta$ is continuous, as in \textsc{glm}s, a typical choice, which will also be used hereafter, is to take Gaussian-like sites $q_{i}(\bbeta)\propto\exp\{-\frac{1}{2}\bbeta^{\intercal}\bQ_{i}\bbeta +\bbeta^{\intercal}\br_{i}\}$ for $i=0,\ldots,n$, with the natural parameters $\br_{i}$ and $\bQ_{i}$ to be determined by the \ep algorithm optimization rules.
As a consequence, the global approximation $q_{\ep}(\bbeta)$ is Gaussian, with $q_{\ep}(\bbeta) = \phi_{p}(\bbeta-\bxi_{\ep},\bOmega_{\ep})$, where $\bxi_{\ep}=\bQ_{\ep}^{-1}\br_{\ep}$ and $\bOmega_{\ep}=\bQ_{\ep}^{-1}$ are obtained from the global natural parameters $\br_{\ep}=\sum_{i=0}^{n} \br_{i}$, $\bQ_{\ep} = \sum_{i=0}^{n}\bQ_{i}$.
If the prior $p(\bbeta)$ has the same form of $q_{0}(\bbeta)$, i.e.\ $p(\bbeta)$ is a multivariate Gaussian density, then we can set $q_{0}(\bbeta)=p(\bbeta)$ and compute only the natural parameters $\br_{i}$ and $\bQ_{i}$ for sites $i=1,\ldots,n$.
It is worth noting that the sites $q_{i}(\bbeta)$ provide an approximation of the likelihood term $p(y_{i}\mid \bbeta)$ and so $q_{i}(\bbeta)$ is not a probability density in $\bbeta$.
Hence, it generally does not hold that $\int q_{i}(\bbeta) d\bbeta=1$.
Likewise, the contributions $\bQ_{i}$'s of each site $i=1,\dots,n$ to the global precision matrix $\bQ_{\ep}$ are not explicitly constrained to be positive definite.
$\bQ_{\ep}$ must instead be positive definite, since the global approximation $q_{\ep}(\bbeta)$, representing the \ep surrogate of the exact posterior distribution, is a valid probability density in $\bbeta$.

In principle, one could think of targeting a global approximation $q_{\ep}(\bbeta)$ minimizing the Kullback-Leibler divergence \citep{kullback1951information} $\textsc{kl}\big[p(\bbeta\mid \by)||q_{\ep}(\bbeta)\big]$.
However, this would involve the computation of integrals with respect to the posterior distribution, which is usually intractable or computationally impractical.
\ep overcomes this issue by iteratively updating the natural parameters $\br_{i}$ and $\bQ_{i}$ of site $i$ minimizing a different \textsc{kl} divergence, informed by the remaining $\{\br_{j}\}_{j\neq i}$ and $\{\bQ_{i}\}_{j\neq i}$, which instead is computationally tractable.
More specifically, at each iteration \ep updates each site $i=1,\ldots,n$ as follows.
Keeping the parameters for all sites $j\ne i$ fixed, first compute the so-called \textit{cavity density}
\begin{equation*}
    q_{-i}(\bbeta)\propto \dfrac{q_{\ep}(\bbeta)}{q_{i}(\bbeta)} \propto \exp \left\{-\frac{1}{2}\bbeta^{\intercal}\bQ_{-i}\bbeta +\bbeta^{\intercal}\br_{-i}\right\}, 
\end{equation*}
where $\br_{-i} = \sum_{j\ne i}\br_{j}$ and $\bQ_{-i} = \sum_{j\ne i}\bQ_{j}$.
Contrary to the local sites, $q_{-i}(\bbeta)$ is a proper distribution in $\bbeta$.
As such, defining $\Psi(\br,\bQ)=\int\exp\left\{-\dfrac{1}{2}\bbeta^{\intercal}\bQ\bbeta +\bbeta^{\intercal} \br\right\}d\bbeta$ so that $\log \Psi(\br,\bQ) = \dfrac{1}{2}\br^{\intercal} \bQ^{-1}\br + \dfrac{p}{2}\log(2\pi) - \dfrac{1}{2}\log |\bQ|$, one gets
\begin{equation*}
    q_{-i}(\bbeta) = \dfrac{1}{\Psi(\br_{-i},\bQ_{-i})} \exp \left\{-\frac{1}{2}\bbeta^{\intercal}\bQ_{-i}\bbeta +\bbeta^{\intercal}\br_{-i} \right\}=\phi_{p}(\bbeta-\bQ_{-i}^{-1}\br_{-i},\bQ_{-i}^{-1}).
\end{equation*}
Similarly, one defines the so-called \textit{hybrid (or tilted) distribution}
\begin{equation*}
    h_{i}(\bbeta)=\dfrac{1}{Z_{h_{i}}} p(y_{i}\mid \bbeta) q_{-i}(\bbeta),
\end{equation*}
with $Z_{h_{i}}=\int p(y_{i}\mid \bbeta) q_{-i}(\bbeta)d\bbeta$ and update the parameters $\br_{i}$ and $\bQ_{i}$ of site $i$ to $\br_{i}^{\new}$ and $\bQ_{i}^{\new}$ so that the resulting global approximation $q_{\ep}^{\new}(\bbeta)\propto q_{i}^{\new}(\bbeta)q_{-i}(\bbeta)$ minimizes $\textsc{kl}\big[h_{i}(\bbeta)||q_{\ep}^{\new}(\bbeta)\big]$.
When the $q_i(\bbeta)$ are taken from an exponential family, the latter minimization is achieved by matching the moments of the sufficient statistics of $q_{\ep}^{\new}(\bbeta)$ computed with respect to $h_{i}(\bbeta)$ and $q_{\ep}^{\new}(\bbeta)$ \citep{seeger2003bayesian,bishop2006pattern}.
Calling $\bmu_{h_{i}} = \E_{h_{i}(\bbeta)} [\bbeta]$ and $\bSigma_{h_{i}} = \Var_{h_{i}(\bbeta)} [\bbeta]$, the \ep moment-matching conditions for site $i$ translate into imposing the equality
\begin{equation*}
	\begin{cases}
	\left(\bQ_{-i}+\bQ_{i}^{\new}\right)^{-1}(\br_{-i} +\br_{i}^{\new}) = \bmu_{h_{i}}\\
	\left(\bQ_{-i}+\bQ_{i}^{\new}\right)^{-1} = \bSigma_{h_{i}},
	\end{cases}
\end{equation*}
which implies
\begin{equation}
\label{eq: momentMatching}
	\begin{cases}
	\br_{i}^{\new} = \left(\bQ_{-i}+\bQ_{i}^{\new}\right) \bmu_{h_{i}} - \br_{-i}\\
	\bQ_{i}^{\new} = \bSigma_{h_{i}}^{-1} - \bQ_{-i}.
	\end{cases}
\end{equation}
At each iteration, the moment-matching conditions \eqref{eq: momentMatching} are enforced sequentially for each site $i=1,\ldots,n$, until convergence is eventually reached.
It is well-known that \ep might be prone to converge issues in some settings \citep{minka2004power}.
When this happens, damping techniques can be applied to reach convergence, although usually at a lower rate \citep{vehtari2020expectation}.
After convergence is reached, one can compute the \ep global approximation $q_{\ep}(\bbeta) = \phi_{p}(\bbeta-\bxi_{\ep},\bOmega_{\ep})$, where $\bxi_{\ep}=\bQ_{\ep}^{-1}\br_{\ep}$ and $\bOmega_{\ep}=\bQ_{\ep}^{-1}$ and $\br_{\ep}=\sum_{i=0}^{n} \br_{i}$, $\bQ_{\ep} = \sum_{i=0}^{n}\bQ_{i}$.

To completely specify the \ep algorithm, initial values of the site parameters need to be provided.
Following common practice employed in the case of multivariate Gaussian prior $p(\bbeta)=\phi_{p}(\bmm,\bC)$ \citep{vehtari2020expectation}, we consider the case in which $q_{\ep}(\bbeta)$ is initialized to the prior distribution, that is we initialize $\br_{0}=\bC^{-1}\bmm$, $\bQ_{0}=\bC^{-1}$, while $\br_{i}=\boldsymbol{0}$, $\bQ_{i}=\boldsymbol{0}$ for $i=1,\ldots,n$.

\subsection{Approximation of the marginal likelihood}
\textsc{ep} can be used as well to obtain an estimation of the marginal likelihood $p(\by)=\int \prod_{i=1}^{n} p(y_{i}\mid \bbeta) p(\bbeta) d\bbeta$ at no additional cost \citep{vehtari2020expectation,minka2001expectation}.
To this end, it becomes necessary to specify a proportionality constant $Z_{i}$ for each site approximation $q_{i}(\bbeta)$, so that
\begin{equation*}
    q_{i}(\bbeta)=\dfrac{1}{Z_{i}}\exp\left\{-\dfrac{1}{2}\bbeta^{\intercal}\bQ_{i}\bbeta +\bbeta^{\intercal} \br_{i}\right\}.
\end{equation*}
Such $Z_{i}$'s were not strictly required to compute the moments of the global posterior approximation $q_{\ep}(\bbeta)$. 
It turns out that the former task is straightforward when the normalizing constant of the hybrid distribution $Z_{h_{i}}$ can be computed in closed form (as is the case for the probit model).
Since $q_{0}(\bbeta)=p(\bbeta)$, it holds $\log Z_{0}=\log\Psi(\br_{0},\bQ_{0})$.
Recall that each $q_{i}(\bbeta)$, for $i=1,\ldots,n$, is not a probability density in $\bbeta$, but it is rather an approximation of the likelihood term $p(y_{i}\mid \bbeta)$.
Adhering to such rigorous representation in terms of global normalization constant $Z_{\ep}$ and local proportionality constants $Z_{i}$, the overall \textsc{ep} approximation can be rewritten as $q_{\ep}^{\new}(\bbeta) = \dfrac{1}{Z_\textsc{ep}^{\new}} q_{i}^{\new}(\bbeta) q_{-i}(\bbeta)$, where, analogously to $Z_{h_{i}}$, one has
\begin{equation*}
\begin{split}
    Z_\textsc{ep}^{\new} & = \int q_{-i}(\bbeta) q_{i}^{\new}(\bbeta) 
 d\bbeta = \dfrac{1}{\Psi(\br_{-i},\bQ_{-i}) \, Z_{i}^{\new}} \int e^{-\frac{1}{2}\bbeta^{\intercal}\bQ_\textsc{ep}^{\new} \bbeta +\bbeta^{\intercal} \br_\textsc{ep}^{\new} } d\bbeta \\
    & = \dfrac{\Psi(\br_\textsc{ep}^{\new},\bQ_\textsc{ep}^{\new})}{\Psi(\br_{-i},\bQ_{-i}) \, Z_{i}^{\new}}.
\end{split}
\end{equation*}
Such reformulation is not immediately useful per se, since the 
the values of these $Z_{i}^{\new}$'s are not determined by the \ep moment-matching conditions.
Accordingly, an additional constraint must be fixed to uniquely identify them, and, in turn, to produce an approximation of the marginal likelihood.
This can be done by extending the moment matching procedure to the so-called zeroth order moments, imposing that $Z_\textsc{ep}^{\new} = Z_{h_{i}}$.
Intuitively, this can be interpreted as a way of enforcing $q_{i}^{\new}(\bbeta)$ to be the best approximation of $p(y_{i}\mid\bbeta)$, such that $\int q_{-i}(\bbeta)q_{i}^{\new}(\bbeta)d\bbeta=\int q_{-i}(\bbeta)p(y_{i}\mid\bbeta)d\bbeta$.
Accordingly, it is trivial to see that
\begin{equation}
\label{eq: updateZi}
    \log Z_{i}^{\new} = \log \Psi(\br_{\ep}^{\new},\bQ_{\ep}^{\new}) - \log \Psi(\br_{-i},\bQ_{-i}) - \log Z_{h_{i}}.
\end{equation}
Upon convergence of the main \textsc{ep} routine, the \ep approximation of the marginal likelihood can be computed by replacing each $p(y_{i}\mid \bbeta)$ with the corresponding approximate site the factor $q_{i}(\bbeta)$, so that
\begin{equation*}
    m_{\ep}(\by) = \int \prod_{i=1}^{n} q_{i}(\bbeta) q_{0}(\bbeta)d\bbeta =
    \dfrac{\int \exp \left\{ {\textstyle -\dfrac{1}{2}\bbeta^{\intercal}\bQ_{\ep}\bbeta +\bbeta^{\intercal} \br_{\ep} } \right\} d\bbeta }{\Psi(\br_{0},\bQ_{0})\prod_{i=1}^{n} Z_{i}} = 
    \dfrac{\Psi(\br_{\ep},\bQ_{\ep})}{\Psi(\br_{0},\bQ_{0})\prod_{i=1}^{n} Z_{i}}.
\end{equation*}
Equivalently, its logarithm reads
\begin{equation}
\label{eq: compML}
    \log m_{\ep}(\by) = \log \Psi(\br_{\ep},\bQ_{\ep}) - \log \Psi(\br_{0},\bQ_{0}) - \sum_{i=1}^{n} \log Z_{i}.
\end{equation}

\section{Efficient updates in GLMs}\label{sec: EPglm}

\subsection[Posterior moments with linear cost in n]{\mbox{Posterior moments and marginal likelihood with linear cost in $n$}}
\label{subsec: post_{n}}

The \textsc{ep} routine presented above applies regardless of the details of the specific $p(y_{i}\mid \bbeta)$ considered.
However, the implementation of such a scheme for \textsc{glm}s greatly benefits from a general structure common to most of them.
This is because the associated likelihood contributions depend on the parameters $\bbeta$ only thorough the inner product $\bx_{i}^{\intercal} \bbeta$ so that $p(y_{i}\mid \bbeta) = \ell_{i}(\bx_{i}^{\intercal} \bbeta)$.
Here the subscript $i$ conveys the dependence of $\ell_{i}$ on the response $y_{i}$ and possibly on additional nuisance parameters $\upsilon_{i}$ of the \textsc{glm}, although this is irrelevant to the structure we want to elicit in the \textsc{ep} approximate (i.e., $\ell_{i}(\bx_{i}^{\intercal} \bbeta)= p(y_{i} \mid \bx_{i}^{\intercal} \bbeta, \upsilon_{i})$).
In the remainder, possible additional parameters $\upsilon_{i}$ are considered fixed constants and thus omitted from the conditioning.

The crucial resulting feature is that the corresponding approximate Gaussian site depends on $\bbeta$ only via $\bx_{i}^{\intercal} \bbeta$ as well
\citep{seeger2007bayesian, seeger2008bayesian}.
This can be proved by studying the moment-generating function (\textsc{mgf}) of the hybrid distribution, as detailed in Appendix~\ref*{appendix: MGF}.
Defining $\bOmega_{i} = \Var_{q_{-i}(\bbeta)} [\bbeta] = \bQ_{-i}^{-1}$ and $\bxi_{i} = \E_{q_{-i}(\bbeta)} [\bbeta] = \bQ_{-i}^{-1} \br_{-i}$, this entails a rank one update of the global \textsc{ep} parameters since the first two moments of the hybrid distribution take the following form.
\begin{Lem}
    \label{lemma: hybridMoments}
    For any model of the form \eqref{eq: BayesianModel} with $p(y_{i}\mid \bbeta) = \ell_{i}(\bx_{i}^{\intercal} \bbeta)$, the moments of the hybrid distribution $h_{i}(\bbeta)$ needed for the \ep updates \eqref{eq: momentMatching} take the form
\begin{equation*}
\begin{split}
    \bmu_{h_{i}} &= \bxi_{i} +  \bOmega_{i}\bx_{i} (\bx_{i}^{\intercal} \bOmega_{i}\bx_{i})^{-1} \big( \E_{h_{i}(\bbeta)} [\bx_{i}^{\intercal}\bbeta] - \bx_{i}^{\intercal} \bxi_{i} \big)  \\
    \bSigma_{h_{i}} &= \bOmega_{i} + \bOmega_{i}\bx_{i} (\bx_{i}^{\intercal} \bOmega_{i}\bx_{i})^{-1} \big( \Var_{h_{i}(\bbeta)}[\bx_{i}^{\intercal}\bbeta] - \bx_{i}^{\intercal} \bOmega_{i}\bx_{i} \big) (\bx_{i}^{\intercal} \bOmega_{i}\bx_{i})^{-1} \bx_{i}^{\intercal} \bOmega_{i} \;.
 \end{split}
\end{equation*}
\end{Lem}
This reduces the dimensionality of the model-dependent moment matching since the latter entails solely the moments induced by the hybrid $h_{i}(\bbeta)$ on the linear subspace identified by the projection on $\bx_{i}$ \citep{zhou2023fast}.
Crucially, this means that the update of the $i$-th site via conditions \eqref{eq: momentMatching} does not require the direct computation of $\bSigma_{h_{i}}^{-1}$.
Indeed, the natural parameters associated with the resulting updated site $q_i(\bbeta)$ are parametrized by scalar quantities, as clarified in the following proposition.
\begin{Prop}
    \label{prop: momentUpdate}
    For any model of the form \eqref{eq: BayesianModel} with $p(y_{i}\mid \bbeta) = \ell_{i}(\bx_{i}^{\intercal} \bbeta)$, the updated parameters of the $i^{th}$ \textsc{ep} site, $i=1,\ldots,n$, have the form $\br_{i}^{\new} = m_{i}^{\new} \bx_{i}$ and $\bQ_{i}^{\new} = k_{i}^{\new} \bx_{i} \bx_{i}^{\intercal}$, with $m_{i}^{\new}$ and $k_{i}^{\new}$ scalar quantities defined as
    \begin{equation*}
    \begin{split}
        k_{i}^{\new} &= \operatorname{var}_{h_{i}(\bbeta)} [\bx_{i}^{\intercal}\bbeta]^{-1} - (\bx_{i}^{\intercal} \bOmega_{i}\bx_{i})^{-1} \\[2pt]
        m_{i}^{\new} &= \operatorname{var}_{h_{i}(\bbeta)} [\bx_{i}^{\intercal}\bbeta]^{-1} \E_{h_{i}(\bbeta)} [\bx_{i}^{\intercal}\bbeta] - (\bx_{i}^{\intercal} \bOmega_{i}\bx_{i})^{-1} \bx_{i}^{\intercal} \bOmega_{i}\br_{-i} \; .
    \end{split}
    \end{equation*}
\end{Prop}
Hence, one can implement \textsc{ep} without storing and updating the $p\times p$ matrices $\bQ_{i}$, but rather storing only scalar quantities $k_{i}$ and $m_{i}$ within
\begin{equation*}
    q_{i}(\bbeta) \propto \exp \Big\{ -\frac{1}{2} k_{i} (\bx_{i}^{\intercal} \bbeta)^{2} + m_{i} (\bx_{i}^{\intercal} \bbeta) \Big\} \; , \qquad \operatorname{for} \; i=1,\ldots,n.
\end{equation*}

As mentioned before, the global approximation is typically initialized to match the prior distribution, which boils down to setting all $k_{i}$'s and $m_{i}$'s to zero.
At each iteration, $k_{i}$ and $m_{i}$ are then updated according to Proposition \ref{prop: momentUpdate}.
The associated updates involve the potentially large matrix inversion $\bOmega_{i}=\bQ_{-i}^{-1}$.
Nonetheless, computational efficiency can be preserved thanks to Woodbury's identity, which gives
\begin{equation*}
 \bOmega_{i}  = \bQ_{-i}^{-1} = \left(\bQ_{\ep} - k_{i} \bx_{i} \bx_{i}^{\intercal} \right)^{-1}
    =  \bOmega_{\ep} + \dfrac{k_{i}}{1- k_{i} \bx_{i}^{\intercal} \bOmega_{\ep} \bx_{i}} \left(\bOmega_{\ep} \bx_{i}\right)\left(\bOmega_{\ep} \bx_{i}\right)^{\intercal},
\end{equation*}
and avoids any explicit $\mathcal{O}(p^3)$ matrix inversion.
Likewise, the updated global covariance $\bOmega_{\ep}^{\new}=(\bQ_{-i}+\bQ_{i}^{\new})^{-1}=\bSigma_{h_{i}}$ can be updated without matrix inversion as in Lemma~\ref{lemma: hybridMoments}.

The same structure can be exploited in the computation of the approximate marginal likelihood, presented in Section \ref{sec: EP}, while matching the same reduction in computational costs.
To this end, the proportionality constant $Z_{i}$ of each site needs to be updated at each iteration together with the natural parameters $\br_{i}$ and $\bQ_{i}$ according to the rule \eqref{eq: updateZi}.
In the case of \textsc{glm}s, some simplifications occur also in this respect.
The computation of the normalizing constant for the hybrid distribution benefits as well of the dimensionality reduction, since now $ Z_{h_{i}}= \int \ell_{i}(\eta_{i}) \phi \big(\eta_{i} - \bx_{i}^{\intercal} \bOmega_{i} \br_{-i}, \, \bx_{i}^{\intercal} \bOmega_{i} \bx_{i} \big)d\eta_{i} $.
Moreover, the structure of $\br_\textsc{ep}$ and $\bQ_\textsc{ep}$ can be further leveraged, leading to the update reported in the proposition below.
\begin{Prop}
    \label{prop: updateZi}
    For any model of the form \eqref{eq: BayesianModel} with $p(y_{i}\mid \bbeta) = \ell_{i}(\bx_{i}^{\intercal} \bbeta)$,
    the updated value of the log-proportionality constant for the $i$-th \ep site, $\log Z_{i}$, $i=1,\ldots,n$, equals
    \begin{equation*}
    \begin{split}
        \log Z_{i}^{\new} = \dfrac{1}{2}\bigg[
        &\dfrac{1}{1+k_{i}^{\new}\bx_{i}^{\intercal} \bOmega_{i}\bx_{i}}\left(2m_{i}^{\new} \br_{-i}^{\intercal} \bOmega_{i} \bx_{i}+(m_{i}^{\new})^{2} \bx_{i}^{\intercal} \bOmega_{i}\bx_{i} -k_{i}^{\new} (\br_{-i}^{\intercal} \bOmega_{i} \bx_{i})^{2} \right)\\
        &-\log (1+k_{i}^{\new} \bx_{i}^{\intercal} \bOmega_{i} \bx_{i}) \bigg]
        -\log Z_{h_{i}}.
    \end{split}
    \end{equation*}
\end{Prop}
After the algorithm has converged, one can compute the final \ep approximate marginal likelihood thanks to \eqref{eq: compML}, which under a spherical Gaussian prior $p(\bbeta) = \phi_{p}(\bbeta,\nu^{2} \bI_{p})$ simplifies to
\begin{equation*}
    \log m_{\ep}(\by) = \dfrac{1}{2} \big[ \,
    \br_{\ep}^{\intercal}\bxi_{\ep} - \log|\bQ_{\ep}| - 2p\log\nu \, \big] - \sum_{i=1}^{n} \log Z_{i}.
\end{equation*}
It is worth noting that the $\mathcal{O}(p^3)$ direct computation of $\log|\bQ_{\ep}|$ can be avoided.
After initializing $\log|\bQ_{\ep}|$ to $\log|\bQ_{0}|=-2p\log\nu$,
the log-determinant can be updated coherently with the moment-matching step time of each site $i$.
Indeed, one has
\begin{equation}
\label{eq: updateLogDetQ}
\begin{split}
    \log|\bQ_{\ep}^{\new}| &= \log \big| \bQ_{\ep} \big[ \bI_{p}+(k_{i}^{\new}-k_{i})\bOmega_{\ep}\bx_{i}\bx_{i}^{\intercal} \big] \big|\\
    &=\log \big|\bQ_{\ep} \big| + \log \big[ 1+(k_{i}^{\new}-k_{i}) \bx_{i}^{\intercal} \bOmega_{\ep}\bx_{i} \big].
\end{split}
\end{equation}

\begin{algorithm}[ht]
 \caption{\textsc{ep glm} - $\mathcal{O}(p^{2} n)$ cost per iteration}
 \label{algo1}
\kwInit{\\
$\bOmega_{\ep} = \nu^{2} \bI_{p};\,\ \br_{\ep}=\boldsymbol{0}; \,\ \log|\bQ_{\ep}|=-2p\log\nu;$\\
$k_{i} = 0$, $m_{i} = 0$, and $\log Z_{i}=0$ for $i=1,\ldots,n$.}
\vspace*{0.1cm}
\kwOpt{}
\vspace*{0.05cm}
\For{$\, t \,$ from $\, 1 \,$ until convergence $ $}{
\For{$\, i \,$ from $\, 1 \,$ to $\, n \,$}{
 \textcolor{gray}{\dotfill} \textcolor{gray}{\small Cavity distribution} \\[1.5pt]
 $\bOmega_{i} = \bOmega_{\ep} + k_{i}/\left(1- k_{i} \bx_{i}^{\intercal} \bOmega_{\ep} \bx_{i}\right) \left(\bOmega_{\ep} \bx_{i}\right)\left(\bOmega_{\ep} \bx_{i}\right)^{\intercal} $ \\[2pt]
 $\br_{-i} = \br_{\ep} - m_{i} \bx_{i} $\\[-4pt]
 \textcolor{gray}{\dotfill} \textcolor{gray}{\small Hybrid distribution} \\[1.5pt]
 $Z_{h_{i}}= \int \ell_{i}(\eta_{i}) \phi \big(\eta_{i} - \bx_{i}^{\intercal} \bOmega_{i} \br_{-i}, \, \bx_{i}^{\intercal} \bOmega_{i} \bx_{i} \big)d\eta_{i} $ \\[2pt]
 $\mu_{i} = \E_{h_{i}(\bbeta)} [\bx_{i}^{\intercal}\bbeta]$ \\[2pt]
 $\rho_{i}^{2} = \operatorname{var}_{h_{i}(\bbeta)} [\bx_{i}^{\intercal}\bbeta] $\\[-3pt]
 \textcolor{gray}{\dotfill} \textcolor{gray}{\small $i$-th site approximation} \\[1.5pt]
 $k_{i}= 1 \,/\, \rho_{i}^{2} - 1 \,/\, \bx_{i}^{\intercal} \bOmega_{i}\bx_{i}  $\\[2pt]
 $m_{i} = \mu_{i} \,/\, \rho_{i}^{2} - \bx_{i}^{\intercal} \bOmega_{i}\br_{-i} \,/\, \bx_{i}^{\intercal} \bOmega_{i}\bx_{i} $\\[3pt]
 \mbox{$\log Z_{i} = \dfrac{1}{2}\bigg[ \dfrac{2m_{i} \br_{-i}^{\intercal} \bOmega_{i} \bx_{i}+m_{i}^{2} \bx_{i}^{\intercal} \bOmega_{i}\bx_{i} -k_{i} (\br_{-i}^{\intercal} \bOmega_{i} \bx_{i})^{2}}{1+k_{i}\bx_{i}^{\intercal} \bOmega_{i}\bx_{i}} - \log (1+k_{i} \bx_{i}^{\intercal} \bOmega_{i} \bx_{i}) \bigg] $ }\\[3pt]
        \hspace*{1.2cm} $-\log Z_{h_{i}}$\\[-4pt]
 \textcolor{gray}{\dotfill} \textcolor{gray}{\small Global approximation} \\
 $\br_{\ep} = \br_{-i} + m_{i} \bx_{i} $\\[2pt]
 $\bOmega_{\ep}=\bOmega_{i} + \rho_{i}^{2}(\bOmega_{i}\bx_{i})(\bOmega_{i}\bx_{i})^{\intercal}$\\[2pt]
 $\log|\bQ_{\ep}| = \log|\bQ_{\ep}| + \log[1+\delta_{k_{i}} \bx_{i}^{\intercal} \bOmega_{\ep}\bx_{i}]$
}
}
\vspace*{4pt}\kwComp{}
$\bxi_{\ep}=\bOmega_{\ep} \br_{\ep}$ \\[2pt]
$\log m_{\ep}(\by) = \tfrac{1}{2} \big( \br_{\ep}^{\intercal}\bxi_{\ep} - \log|\bQ_{\ep}| - 2p\log\nu \big) - \sum_{i=1}^{n} \log Z_{i}$ \\
\vspace*{2pt}\KwOut{$(\bxi_{\ep},\bOmega_{\ep},\log m_{\ep}(\by))$}
\end{algorithm}

Combining all the structural reformulations presented in this section one gets Algorithm~\ref{algo1}.
Except for the computation of the approximate marginal likelihoods, Algorithm~\ref{algo1} extends to the general \textsc{glm}s framework the \textsc{ep} algorithm presented in \cite{chopin2017leave} and implemented in the \texttt{R} package \texttt{EPGLM} for probit and logit regressions.
However, the exploitation of simplifications and iterative updates following from Propositions \ref{prop: momentUpdate}--\ref{prop: updateZi} and equation \eqref{eq: updateLogDetQ} leads to substantial computational advantages (see Section \ref{sec: illustration}).
Note that if the interest is only on posterior moments approximation, the $Z_{i}$ are not needed, and neither $|\bQ_{\ep}|$ is, so in such a case the algorithm can be implemented avoiding the corresponding lines.
Noticeably, Algorithm~\ref{algo1} has per-iteration algebraic cost of $\mathcal{O}(p^{2}n)$, since the update of each site $i=1,\ldots,n$ requires the computation of quadratic forms involving $p\times p$ matrices and the computation of the product between $p\times p$ matrices and $p$-dimensional vectors.
This has to be complemented with a further $\mathcal{O}(n \, \mathcal{C}_\textsc{mm})$ cost per iteration, where $\mathcal{C}_\textsc{mm}$ represent the cost of each model-dependent moment matching computations.
The implementation from Algorithm~\ref{algo1}
is efficient as long as $p$ is of the order of a few hundred, and in particular when $p<n$.
However, although explicit $p\times p$ matrix inversions are avoided, this computational cost might be computationally inefficient or impractical in high-dimensional settings.
To overcome this issue, in Section \ref{subsec: post_p} we derive in full detail a general implementation of \textsc{ep} for Bayesian \textsc{glm}s \eqref{eq: BayesianModel} having per-iteration cost~$\mathcal{O}(pn^{2})$.
The key to achieving such improved scaling lies in performing a simultaneous update of sufficient functionals of the global covariance $\bQ_{\textsc{ep}}$ and all local ones $\bQ_{i}$ at every site update, as first noted in \cite{anceschi2023bayesian} for a specific class of models generalizing the probit regression.

\FloatBarrier

\subsection[Posterior moments with linear cost in p]{\mbox{Posterior moments and marginal likelihood with linear cost in $p$}}
\label{subsec: post_p}

A close inspection of Algorithm \ref{algo1} shows that the $p\times p$ dimensional matrices $\bOmega_{\ep}$ and $\bOmega_{i}$ needed for the optimization of the $i$-th site enter into the computations always by multiplication with the vector $\bx_{i}$.
This includes both the algebraic component of updates and the computation of the reduced hybrid moments.
As a consequence, the whole \textsc{ep} routine can be written directly in terms of the $p$-dimensional vectors $\bw_{i} = \bOmega_{i} \bx_{i}$ and $\bv_{i} = \bOmega_{\ep}\bx_{i}$, $i=1,\ldots,n$.
In turn, such quantity can be efficiently computed and updated in a vectorized form, as shown in Proposition \ref{prop: updatesLargeP}.
Notice that, at each iteration, updating the value of $k_{i}$ to  $k_{i}^{\new}$ for any $i$ induces a new value of $\bQ_{i}$ and, as a consequence, of $\bOmega_{\ep}$.
This means that each time a site $i$ is updated, one has to recompute the corresponding new values of all the $\bv_{j}$'s, $j=1,\ldots,n$, as reported in Proposition \ref{prop: updatesLargeP}, whose proof can be found in the appendix.
\begin{Prop}
\label{prop: updatesLargeP}
For any model of the form \eqref{eq: BayesianModel} with $p(y_{i}\mid \bbeta) = \ell_{i}(\bx_{i}^{\intercal} \bbeta)$, and calling $\bw_{i} = \bOmega_{i} \bx_{i}$ and $\bv_{i} = \bOmega_{\ep}\bx_{i}$, $i=1,\ldots,n$, it holds
\begin{equation*}
	\bw_{i} = d_{i} \bv_{i},
\end{equation*}
where $d_{i}=(1-k_{i}\bx_{i}^{\intercal}\bv_{i})^{-1}$ and $k_{i}$ refers to the value before site $i$ is updated.
After site $i$ is updated, it holds that, for $j=1,\ldots,n$,
\begin{equation*}
\bv_{j}^{\new} = \bv_{j} - c_{i} (\bx_{i}^{\intercal}\bv_{j}) \bv_{i},
\end{equation*}
where we introduced $c_{i}=(k_{i}^{\new} - k_{i})/ (1 + (k_{i}^{\new} - k_{i})\bx_{i}^{\intercal}\bv_{i})$.
Defining the $p \times n$ matrix $ \bV=[\bv_{1},\bv_2,\dots,\bv_{n}] = \bOmega_{\ep} \bX^{\intercal}$, with $\bX=(\bx_{1},\ldots,\bx_{n})^{\intercal}$,
this update can be written in matrix form as
\begin{equation}
    \label{eq: updateV}
    \bV^{\new} = \bV - c_{i} \bv_{i} \bx_{i}^{\intercal} \bV.
\end{equation}
\end{Prop}
Update \eqref{eq: updateV} is the most expensive computation performed at each site update, having cost $\mathcal{O}(p n)$.
As a consequence, this alternative \ep implementation has per-iteration algebraic-component cost $\mathcal{O}(p n^{2})$, contrary to the $\mathcal{O}(p^{2} n)$ of Algorithm~\ref{algo1}.
This has great practical relevance in large-$p$ settings, as it leads to massive computational gains over routinely implemented functions, see Section \ref{sec: illustration}.
In particular, the linear cost in $p$ makes \ep feasible in settings with $p\gg 1000$, moving forward the frontier of scenarios in which \ep can be effectively used and thus relaxing one of the critical problems mentioned in Section 7.3 of \citet{chopin2017leave}.
\\
The initialization of the newly-introduced quantities  
is trivial when $\bOmega_{\ep}$ is initialized to the prior covariance matrix $\bQ_{0}^{-1}$, as standard practice.
Considering the spherical Gaussian prior case, where $\bQ_{0}^{-1}=\nu^{2}\bI_{p}$, this translates into simply initializing $\bV$ to $\nu^{2}\bX^{\intercal}$.
Furthermore, it is sufficient to specify the initialization and updates only for $\bV$, since each $\bw_{i}$ is proportional to the $i$-th column $\bv_{i}$ of $\bV$.

Contrarily to Algorithm~\ref{algo1}, the global covariance matrix $\bOmega_{\ep}$ is not readily available once the procedure has converged.
Nonetheless, it can still obtained efficiently as 
\begin{equation}\label{eq: Q inv}
      \bOmega_{\ep}=\nu^{2}\bI_{p}-\nu^{2}\bV\bK\bX,
\end{equation}
leading to a post-processing cost of $\mathcal{O}(p^{2} n)$.
Indeed, defined $\bK = \diag(k_{1},\ldots,k_{n})$, one has $\bOmega_{\ep}^{-1}=\bQ_{\ep} = \bQ_{0} + \sum_{i=1}^{n} k_{i}\bx_{i} \bx_{i}^{\intercal} = \nu^{-2} \bI_{p} + \bX^{\intercal} \bK \bX$.
Calling $\bLambda = (\bI_{n}+ \nu^{2}\bK\bX \bX^{\intercal})^{-1}$, Woodbury's identity applied to the former gives $\bOmega_{\ep}=\bQ_{\ep}^{-1}=\nu^{2} \bI_{p} - \nu^{4} \bX^{\intercal} \bLambda \bK \bX $.
This in turn allows to elicit
\begin{equation*}
    \bV = \bOmega_{\ep} \bX^{\intercal} = \nu^{2} \bX^{\intercal}\big[\bI_{n} - \nu^{2} \bLambda\bK \bX \bX^{\intercal}\big]=\nu^{2} \bX^{\intercal}\bLambda\big[\bLambda^{-1} - \nu^{2} \bK \bX \bX^{\intercal}\big]=\nu^{2} \bX^{\intercal}\bLambda.
\end{equation*}
If the interest is only in marginal posterior means and variances, the post-processing cost can be further reduced to $\mathcal{O}(p n)$ by computing only the diagonal in \eqref{eq: Q inv}.
The whole routine is summarized in Algorithm \ref{algo2}, which is then well-suited for large-$p$ settings, and in particular cases where $p>n$.

\begin{algorithm}[ht]
    \caption{Efficient \textsc{ep glm} for large $p$ - $\mathcal{O}(p n^{2})$ cost per iteration\label{algo2}}
 \kwInit{\\
 $\br_{\ep}=\boldsymbol{0};\,\ \bV=\left[\bv_{1},\ldots,\bv_{n} \right]=\nu^{2}\bX^{\intercal}; \,\ \log|\bQ_{\ep}|=-2p\log\nu;$\\
 $k_{i} = 0$, $m_{i} = 0$, and $\log Z_{i}=0$ for $i=1,\ldots,n$.} 
 \vspace*{0.1cm}
\kwOpt{}
\vspace*{0.05cm}
 \For{$\, t \,$ from $\, 1 \,$ until convergence $ $}{
 \For{$\, i \,$ from $\, 1 \,$ to $\, n\,$}{
 \textcolor{gray}{\dotfill} \textcolor{gray}{\small Cavity distribution} \\[1.5pt]
 $\bw_{i} = (1-k_{i} \bx_{i}^{\intercal} \bv_{i})^{-1} \bv_{i} $ \\[2pt]
 $\br_{-i} = \br_{\ep} - m_{i} \bx_{i} $\\[-4pt]
 \textcolor{gray}{\dotfill} \textcolor{gray}{\small Hybrid distribution} \\[1.5pt]
 $Z_{h_{i}}= \int \ell_{i}(\eta_{i}) \phi \big(\eta_{i} - \bx_{i}^{\intercal} \bOmega_{i} \br_{-i}, \, \bx_{i}^{\intercal} \bOmega_{i} \bx_{i} \big)d\eta_{i} $ \\[2pt]
 $\mu_{i} = \E_{h_{i}(\bbeta)} [\bx_{i}^{\intercal}\bbeta]$ \\[2pt]
 $\rho_{i}^{2} = \operatorname{var}_{h_{i}(\bbeta)} [\bx_{i}^{\intercal}\bbeta] $\\[-3pt]
 \textcolor{gray}{\dotfill} \textcolor{gray}{\small $i$-th site approximation} \\[1.5pt]
 $\delta_{k_{i}} = \big( 1 \,/\, \rho_{i}^{2} - 1 \,/\, \bx_{i}^{\intercal} \bw_{i} \big) - k_{i} $\\[2pt]
 $k_{i} = \delta_{k_{i}} + k_{i}  $\\[2pt]
 $m_{i} = \mu_{i} \,/\, \rho_{i}^{2} - \br_{-i}^{\intercal} \bw_{i} \,/\, \bx_{i}^{\intercal} \bw_{i} $\\[3pt]
 \mbox{$\log Z_{i} = \dfrac{1}{2}\bigg[
        \dfrac{2m_{i} \br_{-i}^{\intercal} \bw_{i}+m_{i}^{2} \bx_{i}^{\intercal} \bw_{i} -k_{i} (\br_{-i}^{\intercal} \bw_{i})^{2}}{1+k_{i}\bx_{i}^{\intercal} \bw_{i}} -\log (1+k_{i} \bx_{i}^{\intercal} \bw_{i}) \bigg] - \log Z_{h_{i}}$}\\[3pt]
 \textcolor{gray}{\dotfill} \textcolor{gray}{\small Global approximation} \\
 $\br_{\ep} = \br_{-i} + m_{i} \bx_{i} $ \\[2pt]
 $\bV = \bV - \bv_{i} \left[ \delta_{k_{i}}/ \left(1 + \delta_{k_{i}} \bx_{i}^{\intercal} \bv_{i}\right) \right] \bx_{i}^{\intercal} \bV$\\[2pt]
 $\log|\bQ_{\ep}| = \log|\bQ_{\ep}| + \log[1+\delta_{k_{i}} \bx_{i}^{\intercal} \bv_{i}]$
}
}
\vspace*{4pt}\kwComp{}
$\bOmega_{\ep}=\nu^{2}\bI_{p}-\nu^{2}\bV\bK\bX $, with $\bK = \diag(k_{1},\ldots,k_{n})$\\[2pt]
$\bxi_{\ep}=\br_{\ep} - \nu^{2}\bV(\bK\bX \br_{\ep})$\\[2pt]
$\log m_{\ep}(\by) = \tfrac{1}{2} \big( \br_{\ep}^{\intercal}\bxi_{\ep} - \log|\bQ_{\ep}| - 2p\log\nu \big) - \sum_{i=1}^{n} \log Z_{i}$ \\
\vspace*{2pt}\KwOut{$(\bxi_{\ep},\bOmega_{\ep},\log m_{\ep}(\by))$}
\end{algorithm}


\section[Hybrids moments for \textsc{glm}s]{Hybrids moments in GLMs: some relevant cases}\label{sec: EPhybrid}

The efficient structure for the \textsc{ep} updates described above still left unaddressed the computation of the univariate hybrid moments, associated with the specific model of interest.
In the current section, we provide the details of such computations for several notable \textsc{glm}s, including probit, logit, Poisson, and gamma regression.
Even for such fundamental constructions, the required functionals are often unavailable in closed form.
Probit models constitute an eminent exception, in that the hybrid distributions belong to a tractable family of distributions and the exact \textsc{ep} updates are readily available at no additional computational cost
\citep{anceschi2023bayesian, fasano2023efficient_pred, fasano2023efficient}.
\\
For any model lacking these desirable properties, $Z_{h_{i}}$, $\E_{h_{i}(\bbeta)} [\bx_{i}^{\intercal}\bbeta]$ and $ \operatorname{var}_{h_{i}(\bbeta)} [\bx_{i}^{\intercal}\bbeta] $ could be computed via one-dimensional numerical integration, benefiting from the dimensionality reduction induced by the inner product with $\bx_{i}$.
Choosing a large number $\mathcal{T}$ of subintervals for the domain of the desired integral typically provides an accurate approximation of the exact value.
However, this would result in a $\mathcal{O}(\mathcal{T})$ cost $\mathcal{C}_\textsc{mm}$ per each moment-matching step, besides being prone to potential numerical issues.
To preserve the efficient nature of the \textsc{ep} scheme, we here provide model-specific closed-form approximations of the required moments, cutting down $\mathcal{C}_\textsc{mm}$ to $\mathcal{O}(1)$.

In practice, this boils down to supplying an accurate approximation $\widetilde{Z}_{h_{i}}$ of the normalizing constants
\begin{equation*}
    Z_{h_{i}}= \int \ell_{i}(\eta_{i}) \phi \big(\eta_{i} - \lambda_{i}, \, \rho_{i}^{2} \big)d\eta_{i} 
\end{equation*}
and working out its derivative \citep{chopin2017leave}.
Here $ \lambda_{i} = \bx_{i}^{\intercal} \bOmega_{i} \br_{-i} = \bw_{i}^{\intercal} \br_{-i} $ and $\rho_{i}^{2} = \bx_{i}^{\intercal} \bOmega_{i} \bx_{i} = \bw_{i}^{\intercal} \bx_{i} $, depending on which version of the \textsc{ep} updates is being considered.
Indeed, the required moments can be further obtained from the derivatives of $\log Z_{h_{i}}$ \citep{seeger2005expectation} via
\begin{equation}
\label{eq: moments from diff log Z}
\begin{split}
    \E_{h_{i}(\bbeta)} [\bx_{i}^{\intercal}\bbeta] &= \lambda_{i} + \rho_{i}^{2} \frac{\partial }{\partial \lambda_{i}} \log Z_{h_{i}} \\
    \operatorname{var}_{h_{i}(\bbeta)} [\bx_{i}^{\intercal}\bbeta] &= \rho_{i}^{2} + \rho_{i}^{2} \left( 2 \frac{\partial }{\partial \rho_{i}^{2}} \log Z_{h_{i}} - \Big( \frac{\partial }{\partial \lambda_{i}} \log Z_{h_{i}} \Big)^{2} \right) \rho_{i}^{2} \; ,
\end{split}
\end{equation}
as proven in Appendix~\ref*{appendix: comp}.
In turn, the above expressions can be directly combined with the updates from Propostion~\ref*{prop: momentUpdate}.
Defining $\Theta_{i} = \frac{\partial }{\partial \lambda_{i}} \log Z_{h_{i}}$ and $\Delta_{i} = 2 \frac{\partial }{\partial \rho_{i}^{2}} \log Z_{h_{i}} - \Big( \frac{\partial }{\partial \lambda_{i}} \log Z_{h_{i}} \Big)^{2}$, simply algebraic calculations lead to
\begin{equation*}
    k_{i}^{\new} = - \Delta_{i} \big/ \big( 1 + \rho_{i}^{2} \Delta_{i} \big),  
    \qquad \qquad
    m_{i}^{\new} = k_{i}^{\new} \big( \lambda_{i} - \Delta_{i}^{-1} \Theta_{i} \big) \; .
\end{equation*}
Accordingly, any differentiable approximation $\widetilde{Z}_{h_{i}}$ for $Z_{h_{i}}$ naturally induces approximate updates for the natural parameters of the Gaussian \textsc{ep} sites.
The quality of the resulting approximations will be highly specific to the model considered, the nature of $\widetilde{Z}_{h_{i}}$ and the order of magnitude of the parameters $\lambda_{i}$ and $\rho_{i}^{2}$.

\subsection{Probit regression}\label{sec: EPprobit}

Probit regression represents a prominent example of models for which the hybrids belong to a known family of distributions, further allowing for closed-form \textsc{ep} updates.
This can be seen as a relevant special case of a broader class of models considered in \citet{anceschi2023bayesian}, where the \ep algorithm further simplifies, avoiding the need to work with the possibly computationally challenging $\textsc{sun}$ hybrid distributions, as shown below.
In particular, the form of the \ep updates can be obtained by leveraging results on the efficiently tractable multivariate extended skew-normal (\textsc{sn}) random variables (see \cite{azzalini2014skew}).
In probit models, the binary responses $y_{i} \in \{ 0, 1\}$ are modeled as Bernoulli random variables with success probability $\Phi(\bx_{i}^{\intercal} \bbeta)$, for each $i=1,\ldots,n$.
Accordingly, each likelihood contribution takes the form
\begin{equation}\label{eq: probit}
p(y_{i}\mid \bbeta) = \big( \Phi(\bx_{i}^{\intercal} \bbeta) \big)^{y_{i}} \big( 1-  \Phi(\bx_{i}^{\intercal} \bbeta) \big)^{1-y_{i}} = \Phi\big( (2y_{i}-1) \bx_{i}^{\intercal} \bbeta \big) \; .
\end{equation}
This allowed \cite{durante2019conjugate} to first show that the posterior distribution under Gaussian priors belongs to the \textsc{sun} family.
The same rationale can be extended to the analysis of the \textsc{ep} hybrid distributions $h_{i}(\bbeta)$, resulting in the following proposition.
\begin{Prop}
    \label{prop: hybrid probit}
    For the probit likelihood \eqref{eq: probit}, the hybrid density $h_{i}(\bbeta)$ is the density of a multivariate extended skew-normal distribution $\textsc{sn}_{p}(\bxi_{i},\bOmega_{i},\balpha_{i},\tau_{i})$ (see \cite{azzalini2014skew}), with
\begin{equation*}
\begin{split}
    \bomega_{i} &=\left[\text{diag}\left(\bOmega_{i}\right)\right]^{1/2} 
    \hspace*{50pt}
    \bar{\bOmega}_{i} = \bomega_{i}^{-1} \bOmega_{i} \bomega_{i}^{-1} \\
    \balpha_{i} &= (2y_{i}-1)\bomega_{i}\bx_{i}
    \hspace*{55pt}
    \tau_{i} = (2y_{i}-1)(1+\bx_{i}^{\intercal}\bOmega_{i}\bx_{i})^{-1/2}\bx_{i}^{\intercal}\bxi_{i} \; .
\end{split}
\end{equation*}
and $\bxi_{i} = \bQ_{-i}^{-1}\br_{-i}$ and $\bOmega_{i} = \bQ_{-i}^{-1}$, as before.
Its normalizing constant is $Z_{h_{i}}=\Phi(\tau_{i})$, while the updates for one dimensional parameters in the $i$-th approximate site take the form
\begin{equation*}
\begin{split}
    k_{i}^{\new} &= -\zeta_2(\tau_{i})/\big(1 + \bx_{i}^{\intercal}\bOmega_{i}\bx_{i} + \zeta_2(\tau_{i})\bx_{i}^{\intercal}\bOmega_{i}\bx_{i}\big) \\[5pt]
    m_{i}^{\new} &= \zeta_{1}(\tau_{i}) s_{i} + k_{i}^{\new} \bx_{i}^{\intercal} \bOmega_{i} \br_{-i} + k_{i}^{\new} \zeta_{1}(\tau_{i}) s_{i} \bx_{i}^{\intercal} \bOmega_{i}\bx_{i} \; ,
\end{split}
\end{equation*}
where $s_{i}=(2y_{i}-1)(1+\bx_{i}^{\intercal}\bOmega_{i}\bx_{i})^{-1/2}$, $\zeta_{1}(x) = \phi(x)/\Phi(x)$ and $\zeta_2(x)=-\zeta_{1}(x)^{2}-x\zeta_{1}(x)$.
\end{Prop}
All the above holds even under the reformulation from Section~\ref{subsec: post_p}.
This amounts simply to replacing $\bx_{i}^{\intercal} \bOmega_{i}\bx_{i}$ and $\bx_{i}^{\intercal} \bOmega_{i} \br_{-i}$ with $\bw_{i}^{\intercal} \bx_{i}$ and $\bw_{i}^{\intercal} \br_{-i}$, respectively.

\subsubsection{Closed-form predictive probabilities}
\label{subsec: predProbit}

The analytic tractability of the probit carries over to predictions for new observations.
Indeed, the Gaussian form of the global \ep approximation also has a very convenient side effect that the resulting approximate predictive probabilities for new observations admit a simple closed form.
Considering a new observation $y_{\new}$ having covariate vector $\bx_{\new}$, by analogy with the exact predictive probability $\Pr{}[y_{\new}=1\mid \by]  = \E[y_{\new}\mid \by ]= \E_{p(\bbeta\mid \by)}\big[\Phi\big(\bx_{\new}^{\intercal} \bbeta \big)\big]$, one gets the \ep approximation counterpart
\begin{equation}
\label{eq: pred}
    \Pr{}_{\textsc{ep}}[y_{\new}=1\mid \by] = \E_{q_{\ep}(\bbeta)}\big[\Phi\big(\bx_{\new}^{\intercal} \bbeta \big)\big]
    =\Phi\big( \big(1 + u\big)^{-1/2} \bx_{\new}^{\intercal} \bxi_\textsc{ep}\big),\vspace*{-0.2cm}
\end{equation}
where $u=\bx_{\new}^{\intercal} \bOmega_{\textsc{ep}}\bx_{\new}$ and the last equality in \eqref{eq: pred} follows by Lemma 7.1 in \citet{azzalini2014skew}.
The evaluation of \eqref{eq: pred} does not pose any computational difficulties and has a cost that is always dominated by the \ep algorithm, as the only computationally relevant part is the computation of the quadratic form $u$.
This can be computed directly when using Algorithm \ref{algo1} (usually, when $p<n$), as $\bOmega_{\textsc{ep}}$ is directly returned by the algorithm, and $u$ can be computed at cost $\mathcal{O}(p^{2})$.
On the other hand, when using Algorithm \ref{algo2} (usually, when $p>n$), this direct computation can be avoided.
For simplicity, assume once more a spherical Gaussian prior $p(\bbeta)=\phi_{p}(\bbeta,\nu^{2} \bI_{p})$, so that $q_{0}(\bbeta)$ is fixed and with natural parameter $\br_{0}=\boldsymbol{0}$ and $\bQ_{0}=\nu^{-2}\bI_{p}$.
Then, exploiting \eqref{eq: Q inv}, it can be seen that $u=\nu^{2}\left[\bx_{\new}^{\intercal} \bx_{\new} - \big(\bV^{\intercal}\bx_{\new}\right)^{\intercal} \bK \left(\bX\bx_{\new}\right)\big]$, which can be computed at cost $\mathcal{O}(pn)$.
As a consequence of this argument, the closed-form approximation of the exact predictive probability  \eqref{eq: pred} can be efficiently computed at cost $\mathcal{O}(p\cdot\min\{p,n\})$ from the \textsc{ep} parameters.

\subsection{Logistic regression}
Similarly to the probit case, logistic regression describes binary responses $y_{i} \in \{0,1 \}$ as Bernoulli random variables, for each $i=1,\ldots,n$, although now the success probabilities $(1+e^{-\bx_{i}^{\intercal} \bbeta})^{-1}$ are mediated by the logit link.
In such a case, the likelihood terms become
\begin{equation*}
p(y_{i}\mid \bbeta) = \Big( \frac{1}{1+e^{-\bx_{i}^{\intercal} \bbeta}} \Big)^{y_{i}} \Big( \frac{1}{1+e^{\bx_{i}^{\intercal} \bbeta}} \Big)^{1-y_{i}} = \frac{1}{1+e^{- (2 y_{i}-1)\bx_{i}^{\intercal} \bbeta}}  \; .
\end{equation*}
Contrary to the probit case, the hybrid moments for the logit exact sites do not admit closed-form expressions.
Following \citet{chopin2017leave}, we propose a two-fold application of the approximation introduced by \citet{mackay1992evidence}
\begin{equation*}
\begin{split}
     \int \frac{1}{1+e^{- \eta}}  \phi \big(\eta - \lambda, \, \rho^{2} \big) d\eta \approx \int \Phi\left( \eta \sqrt{\pi/8} \right)  \phi \big(\eta - \lambda, \, \rho^{2} \big) d\eta
     &= \Phi\left(\frac{\lambda}{\sqrt{8\pi + \rho^{2}}} \right) \\ 
     &\approx
      \frac{1}{1+e^{- \lambda / \sqrt{1 + \rho^{2} \pi / 8}}}
\end{split}
\end{equation*}
Accordingly, the hybrid normalizing constants can be approximated as
\begin{equation*}
    Z_{h_{i}} \approx \widetilde{Z}_{h_{i}} \coloneq \frac{1}{1+e^{- (2 y_{i}-1) \lambda_{i} / \sqrt{1 + \rho_{i}^{2} \pi / 8}}} \;.
\end{equation*}
Exploiting the fact that $\frac{\partial}{\partial s} \big( (1+e^{-s})^{-1} \big) = (1+e^{-s})^{-1} \big(1 - (1+e^{-s})^{-1} \big)$, one gets
\begin{equation*}
\begin{split}
    \frac{\partial }{\partial \lambda_{i}} \log \widetilde{Z}_{h_{i}} &= \frac{(2y_{i}-1)}{ \sqrt{1 + \rho_{i}^{2} \pi / 8}} \, (1-\widetilde{Z}_{h_{i}}) \\
     \frac{\partial }{\partial \rho_{i}^{2}} \log \widetilde{Z}_{h_{i}} &= - \frac{1}{2} \frac{\pi}{8} \frac{\lambda_{i}}{ \sqrt{1 + \rho_{i}^{2} \pi / 8}} \, \frac{\partial }{\partial \lambda_{i}} \log \widetilde{Z}_{h_{i}} \; .
\end{split}
\end{equation*}
An approximate implementation of the \textsc{ep} moment matching is thus obtained replacing  $Z_{h_{i}}$, $\frac{\partial }{\partial \lambda_{i}} \log Z_{h_{i}}$ and $\frac{\partial }{\partial \rho_{i}^{2}} \log Z_{h_{i}}$ respectively with $\widetilde{Z}_{h_{i}}$, $ \frac{\partial }{\partial \lambda_{i}} \log \widetilde{Z}_{h_{i}}$ and $\frac{\partial }{\partial \rho_{i}^{2}} \log \widetilde{Z}_{h_{i}}$.

\subsection{Poisson regression}

We now move beyond the modeling of binary responses, focusing instead on count data $y_{i} \in \mathbb{N}$.
In particular, we focus on Poisson regression under the logarithmic link.
This corresponds to a likelihood of the form
\begin{equation*}
p(y_{i}\mid \bbeta) = \frac{\big( e^{\bx_{i}^{\intercal} \bbeta} \big)^{y_{i}} e^{-( e^{\bx_{i}^{\intercal} \bbeta} ) }}{y_{i} !} \; ,
\end{equation*}
in turn, associated with hybrid normalizing constants 
\begin{equation}
\label{eq: norm const poisson}
    Z_{h_{i}} = \frac{1}{y_{i} !} \, e^{y_{i} \lambda_{i} + \frac{1}{2} \rho_{i}^{2} y_{i}^{2}}  \int e^{-e^{\eta_{i}}}  \phi \big(\eta_{i} - (\lambda_{i} + \rho_{i}^{2} y_{i}), \, \rho_{i}^{2} \big) d\eta_{i} \; .
\end{equation}
The integral within the expression above does not admit an exact solution.
Indeed, it is easy to see that it coincides with the Laplace transform $\mathcal{L}(s)$ of a log-normal random variable, evaluated at a specific value of its argument $s$.
More details on the connection with the log-normal Laplace transform are provided in Appendix~\ref*{appendix: comp}.
Several dedicated approximations have been developed over the years.
As a default for the \textsc{ep} updates, we propose exploiting the contribution by \citet{asmussen2014laplace}, which gives
\begin{equation*}
    \widetilde{Z}_{h_{i}} = \frac{1}{y_{i} !} \, e^{y_{i} \lambda_{i} + \frac{1}{2} \rho_{i}^{2} y_{i}^{2}} \, \frac{ e^{-\frac{1}{\rho_{i}^{2}}\mathcal{W}\big(\rho_{i}^{2} e^{\lambda_{i} + \rho_{i}^{2} y_{i}} \big) \left( 1 + \frac{1}{2}\mathcal{W}\big(\rho_{i}^{2} e^{\lambda_{i} + \rho_{i}^{2} y_{i}} \big) \right)} }{ \sqrt{1 + \mathcal{W}\big(\rho_{i}^{2} e^{\lambda_{i} + \rho_{i}^{2} y_{i}} \big) } } \; .
\end{equation*}
Here $\mathcal{W}(x)$ the so-called \textit{Lambert W function}, which is defined only implicitly as the solution of $\mathcal{W}(x) e^{\mathcal{W}(x)} = x$, but that can be evaluated efficiently in most programming languages.
As proved in Appendix~\ref*{appendix: comp}, the derivatives of the logarithm of $\widetilde{Z}_{h_{i}}$ read
\begin{equation*}
\begin{split}
   \frac{\partial}{\partial \lambda_{i}} \log \widetilde{Z}_{h_{i}} &= y_{i} - \mathcal{W} \big( \rho_{i}^{2} e^{\lambda_{i} + \rho_{i}^{2} y_{i}} \big)
   \left( \frac{1}{\rho_{i}^{2}} + \frac{1}{2} \Big( 1+ \mathcal{W} \big( \rho_{i}^{2} e^{\lambda_{i} + \rho_{i}^{2} y_{i}} \big) \Big)^{-2} \right) \\
   \frac{\partial}{\partial \rho_{i}^{2}} \log \widetilde{Z}_{h_{i}} &= \frac{y_{i}^{2}}{2} + \frac{1}{\rho_{i}^{2}} \frac{\mathcal{W} \big( \rho_{i}^{2} e^{\lambda_{i} + \rho_{i}^{2} y_{i}} \big)}{\rho_{i}^{2}} \Big( 1 + \frac{1}{2}  \mathcal{W} \big( \rho_{i}^{2} e^{\lambda_{i} + \rho_{i}^{2} y_{i}} \big) \Big) \\
   & \quad + \frac{1+\rho_{i}^{2} y_{i}}{\rho_{i}^{2}} \Big( \frac{\partial}{\partial \lambda_{i}} \log \widetilde{Z}_{h_{i}} - y_{i} \Big) \; .
\end{split}
\end{equation*}
Our empirical insights suggest that the above provides an accurate approximation as long as $y_{i}>0$ or $\lambda_{i}$ is large enough.
Indeed, \cite{asmussen2014laplace} essentially apply Laplace's method to the Laplace transform $\mathcal{L}(s_{i})$ of the log-normal distribution.
Such an approximation is exact only in the limit $s_{i} \rightarrow \infty$.
In the context of \textsc{ep} Poisson regression, $\mathcal{L}(s_{i})$ needs to be evaluated at $s_{i}=\exp{(\lambda_{i} + \rho_{i}^{2} y_{i})}$ for the $i^{th}$ site.
Observations with $y_{i}=0$ and small $\lambda_{i}$ clearly fall outside the region of large $s_i$.
To preserve accuracy in all settings, we propose employing the alternative approximation scheme by \citet{rossberg2008accurate,rossberg2008laplace} for sites with $y_{i}=0$ and $\lambda_{i}$ smaller than some heuristic threshold.
The details on the resulting approximate moment matching are provided in Appendix~\ref*{appendix: comp}.

\subsubsection{Closed-form predictive means}
\label{subsec: predPoisson}
The combination of the logarithmic link with the Gaussian \ep posterior approximation also brings a very convenient closed-form expression for the predictive mean of a new observation $y_{\new}$ having covariate vector $\bx_{\new}$, that is $\E[y_{\new}\mid \by ]$.
Note that this provides the analog of the closed-form predictive probabilities derived for the probit model in Section \ref{subsec: predProbit} as, in that case, $\Pr{}[y_{\new}=1\mid \by]  = \E[y_{\new}\mid \by ]$.
When a logarithmic link is employed, as in the default choice for the Poisson model (although the same argument holds true for any \textsc{glm}s with a logarithmic link), one has
\begin{equation}
\label{eq: predMeanPoisson}
\E[y_{\new}\mid\by] = \E_{p(\bbeta\mid\by)}\left[\E_{p(y_{\new}\mid\bbeta)}[y_{\new}]
\right]=\E_{p(\bbeta\mid\by)}\left[e^{\bx_{\new}^{\intercal}\bbeta}
\right].
\end{equation}
Thus, by replacing the true posterior distribution with its \ep approximation in \eqref{eq: predMeanPoisson}, the \ep approximate predictive mean for observation $y_{\new}$ is
$
\E_{\ep}[y_{\new}\mid\by] = \E_{q_{\ep}(\bbeta)}\left[e^{\bx_{\new}^{\intercal}\bbeta}
\right].
$
Now, calling $\eta=\bx_{\new}^{\intercal}\bbeta$, since $q_{\ep}(\bbeta)=\phi_{p}(\bbeta-\xi_{\ep},\bOmega_{\ep})$, one has that, according to the \ep approximation of the posterior distribution of $\bbeta$, $\eta\sim N(m_{\new},s_{\new}^{2})$, with $m_{\new}=\bx_{\new}^{\intercal}\bxi_{\ep}$ and $s_{\new}^{2}=\bx_{\new}^{\intercal}\bOmega_{\ep}\bx_{\new}$.
Consequently, $e^\eta$ has log-normal distribution with parameters $m_{\new}$ and $s_{\new}^{2}$ and we immediately obtain
\begin{equation*}
    \E_{\ep}[y_{\new}\mid\by] = \E_{\eta\sim N(m_{\new},s_{\new}^{2})}\left[e^\eta\right]=\exp\{m_{\new}+s_{\new}^{2}/2\},
\end{equation*}
where the quadratic form $\bx_{\new}^{\intercal}\bOmega_{\ep}\bx_{\new}$ can be efficiently computed at cost $\mathcal{O}(p\cdot \min{p,n})$, as explained in Section \ref{subsec: predProbit}, without affecting the overall cost of the \ep procedure.

\subsection{Gamma regression}
Gamma regression is widely used \textsc{glm} to model continuous, non-negative, and positively skewed data.
As the name points out, it postulates gamma distributed responses $y_{i} \mid \bbeta \overset{ind}{\sim} \mathcal{G}a (\upsilon_{i}, \varrho(\bx_{i}^{\intercal} \bbeta) / \upsilon_{i})$.
Here we focus on the logarithmic link $\varrho(\bx_{i}^{\intercal} \bbeta)=e^{\bx_{i}^{\intercal} \bbeta}$ since it is the standard choice in the gamma likelihood \textsc{glm}, while we will consider the additional shape parameter $\upsilon_{i}>0$ to be fixed.
This leads to a likelihood of the form
\begin{equation*}
p(y_{i}\mid \bbeta) = \frac{1}{\Gamma(\upsilon_{i})} \Big( \frac{\upsilon_{i}}{e^{\bx_{i}^{\intercal} \bbeta}}\Big)^{\upsilon_{i}} y_{i}^{\upsilon_{i}-1} e^{-\upsilon_{i} y_{i} e^{-\bx_{i}^{\intercal} \bbeta}} \; ,
\end{equation*}
reflected in hybrid normalizing constants
\begin{equation*}
    Z_{h_{i}} = \frac{\upsilon_{i}^{\upsilon_{i}}}{\Gamma(\upsilon_{i})} \, y_{i}^{\upsilon_{i}-1} \, e^{-\upsilon_{i} \lambda_{i} + \frac{1}{2} \rho_{i}^{2} \upsilon_{i}^{2}} \int  e^{-\upsilon_{i} y_{i} e^{\eta_{i}}} \phi \big(\eta_{i} - (\rho_{i}^{2} \upsilon_{i} -\lambda_{i}), \rho_{i}^{2} \big) d\eta_{i} \; .
\end{equation*}
Similarly to  Poisson regression, the required integral takes the form of the log-normal Laplace transform.
Accordingly, the same heuristics exploited before can be adapted to approximate moment matching in gamma regression as well.

\section{Simulation Studies}\label{sec: simstudies}
\subsection{Probit regression}\label{subsec: simProbit}
In this section, we demonstrate the advantages of our proposed efficient \textsc{ep} through a simulation study.
We apply the probit regression model \eqref{eq: probit}, with $\nu^{2}=25$, to multiple simulated datasets, with $n = 500$ and $p = 125, \, 250, \, 500, \, 1000$ and $2000$.
The $p-1$ covariates, excluding the intercept, are generated independently from a standard normal distribution and are then standardized to have zero mean and $0.5$ standard deviation, following \citet{gelman2008weakly} as in \citet{chopin2017leave}.
Our investigation focuses on the performance of \textsc{ep} when utilizing the efficient implementations outlined in Algorithm~\ref{algo1} and Algorithm~\ref{algo2}, choosing automatically the optimal one based on the dimensionality of the input data.
These implementations are used to compute posterior moments and predictive probabilities (for $\tilde{n}=100$ test units) when $p<n$ and $p\ge n$, respectively.
We refer to this implementation as \textsc{ep-eff} in the following discussion.

We compare the \textsc{ep-eff} approach with the \textsc{pfm-vb} approximation introduced in \cite{fasano2022scalable}, which has higher approximation accuracy than the mean-field variational approach presented in \citet{consonni2007meanfield}.
This comparison is made in terms of both running time and the quality of the approximation.
We measure the latter in terms of the goodness of the approximation of the $p$ posterior means and standard deviations and the $\tilde{n}$ predictive probabilities.
For each of these three quantities, we compute the median absolute difference (together with the first and third quartile) with respect to those computed via $10000$ \textsc{mcmc} samples obtained using the \texttt{rstan} \citep{stan2023rstan} implementation of the No-U-Turn sampler (\textsc{nuts}), which performs Hamiltonian Monte Carlo (\textsc{hmc}) sampling.
These samples were generated by $5$ chains of $2000$ samples each, after a burn-in of $1000$.
For more information on the \textsc{hmc} sampler in probit regression, please refer to \citet{chopin2017leave}.

\begin{figure}
    \centering
    \includegraphics[width=\linewidth,height=5cm]{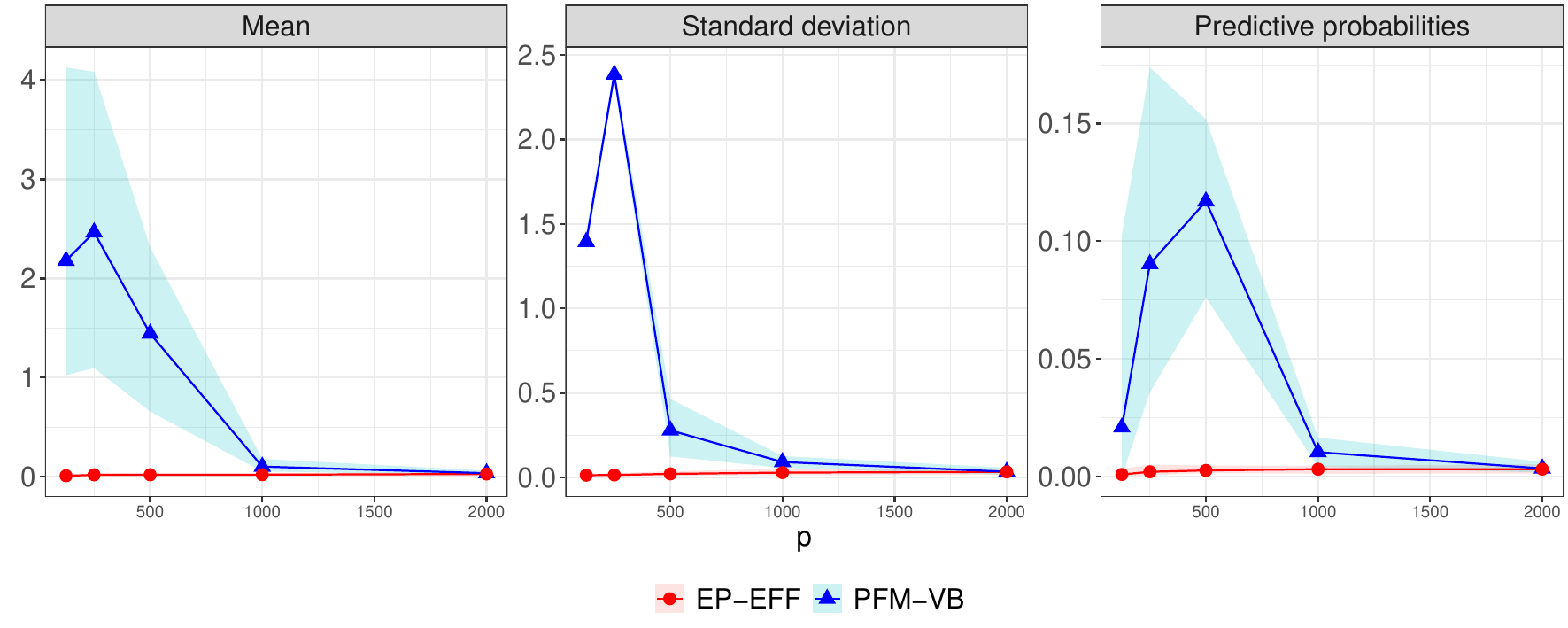}
    \caption{For varying $p$, median absolute difference between the $p$ posterior means, the standard deviations and the $\tilde{n}=100$ predictive probabilities resulting from $10 000$ \textsc{hmc} (\textsc{nuts}) samples and the ones arising from \textsc{ep-eff} and \textsc{pfm-vb} for probit regression with $n=500$ and $\nu^{2}=25$.
    Shaded areas denote the first and third quartiles.
    }
    \label{fig:1}
\end{figure}

As presented in Table~\ref{table: simProbit}, the \textsc{nuts} method is considerably slower, even though running five shorter chains in parallel.
This is why we considered $p$ only up to $2000$ in order to be able to get posterior samples from the \textsc{nuts} method, which is used as a benchmark.
To illustrate the computational advantages in comparison to standard \textsc{ep} implementations, we also compare the running time needed to obtain the \textsc{ep} approximations using the \texttt{R} function \texttt{EPprobit} from the \texttt{EPGLM} package, which implements the \textsc{ep} derivations reported in \cite{chopin2017leave}.
As shown in Table~\ref{table: simProbit}, \textsc{ep-eff} dramatically reduces the computational effort to approximate the posterior moments and predictive probabilities, compared to the standard \texttt{EPprobit} in high dimensions.
This results in a decrease in running time by more than three orders of magnitude when $p=2000$, with computational gains increasing as expected with higher values of $p$.
The \textsc{ep-eff} running times, although generally much lower than the ones of \texttt{EPprobit}, are still higher than the ones of \textsc{pfm-vb} for $p>n$.
For $p<n$ instead, the lower number of iterations needed to reach convergence by \textsc{ep} make \textsc{ep-eff} faster than \textsc{pfm-vb}.
\begin{table*}[t]
\centering
\caption{\footnotesize{Running time, in seconds, to compute posterior moments and predictive probabilities for $\tilde{n}=100$ test observations with the \textsc{ep} approximation as in Algorithms \ref{algo1} and \ref{algo2} (\textsc{ep-eff}), with the \textsc{ep} approximation computed via the \texttt{R} function \texttt{EPprobit} (\texttt{EPprobit}), with the \textsc{pfm-vb} approximation (\textsc{pfm-vb}) and with the \texttt{rstan} implementation of the No-U-Turn \textsc{hmc} sampler (\textsc{nuts}) for probit regression with $n=500$ and $\nu^{2}=25$.
}}
\label{table: simProbit}
\begin{tabular}[c]{ll|ccccc}
 \multicolumn{2}{l|}{}  &  \multicolumn{5}{c}{\textit{p}}   \\
\hline
 & \textit{Method} & 125 & 250 & 500 & 1000 & 2000 \\ 
   \hline
Running time &\ \textsc{ep-eff} &\ 0.55 &\ 1.13 &\ 2.06 &\ 3.88 &\ 8.15 \\
(seconds) &\ \texttt{EPprobit}\ &\ 6.06 &\ 38.36 &\ 232.92 &\ 1867.35 &\ 21206.49 \\
&\ \textsc{pfm-vb}\ &\ 2.88 &\ 1.90 &\ 3.25 &\ 1.21 &\ 1.81 \\
&\ \textsc{nuts}\ &\ 152.13 &\ 491.08 &\  1971.50 &\  7745.48 &\ 20415.65 \\
\hline
\end{tabular}
\end{table*}
Nevertheless, when examining the quality of the approximation of the two posterior moments and predictive probabilities (Figure~\ref{fig:1}), \textsc{ep-eff} consistently provides accurate approximations across different dimensions of $p$, while \textsc{pfm-vb} gets similar accuracy only for $p \gtrsim 2n$.
Combined with the improved accuracy of \textsc{ep} in all scenarios, 
this highlights the importance of the proposed efficient implementations for \textsc{ep}, making it computationally feasible in almost every practical scenario including the challenging high-dimensional and/or large sample-size settings where previously available implementations are either impractical or inaccurate.

Finally, evaluation of the \textsc{ep} approximate predictive probabilities is a direct by-product of the computation of optimal posterior approximation, thanks to the closed-form expression derived in Section \ref{subsec: predProbit}.
Accordingly, the associate additional computation time is minimal in the case of \textsc{ep}.
In contrast, the equivalent step in \textsc{pfm-vb} requires a (slightly) higher cost, due to the lack of a closed-form expression of the predictive probabilities, which are computed from Equation (11) in \citet{fasano2022scalable} sampling from $n$ independent univariate truncated normals.
Although this can be done efficiently and requires no more than one second in the considered scenarios, the availability of a closed form for \textsc{ep} makes this step computationally faster.
For the sake of completeness, we present the running times for the various methods in calculating only the posterior moments in Table \ref*{tab: simProbitOnlyMoments} of the Appendix.
The comparison mirrors that of the running times for predictive probabilities in Table \ref{table: simProbit}.
All computations are carried out in \texttt{R}, using plain \texttt{R} code, on an eight-core i9 a Lenovo ThinkStation P330 with 16Gb RAM (Windows 11 Pro).

\FloatBarrier

\subsection{Poisson regression}
\label{subsec: simPoisson}
We now show the performance of the proposed \ep efficient algorithm for Bayesian Poisson regression under Gaussian prior.
This constitutes a widespread popular Bayesian \textsc{glm}, for which, however, many open challenges remain open due to the lack of a tractable posterior distribution and possible mixing issues in the available sampling methods (see also \citet{dangelo2023efficient}).
We generated synthetic datasets with the same dimensions used in Section \ref{subsec: simProbit} so that sampling methods could also be run and used as a benchmark.
Specifically, we generated five datasets fixing the number of observations $n=500$ and held-out units $\tilde{n}=100$, while we varied $p$ over the values $125, 250, 500, 1000, 2000$.
The generation process for such datasets is as follows.
The $p-1$ covariate values for each observation were preliminarily generated independently from a standard Gaussian.
Then, for each variable, we centered the generated covariate values by subtracting the variable empirical mean and rescaled them by dividing by $2\sqrt{p-1}\widehat{sd}_{j}$, where $\widehat{sd}_{j}$ is the estimated standard deviation for variable $j$.
This is done to have a constant variance of the linear predictor generating the data across the multiple scenarios, since the parameters $\beta_{j}$, $j=2,\ldots,p$ are uniformly sampled from the interval $[-5,5]$, while $\beta_1$ was fixed to $5$.
The model was completed by taking a dimension-adjusted prior on the parameters $\nu^{2}=250/p$.

\begin{figure}[tb]
    \centering
    \includegraphics[width=\linewidth,height=5cm]{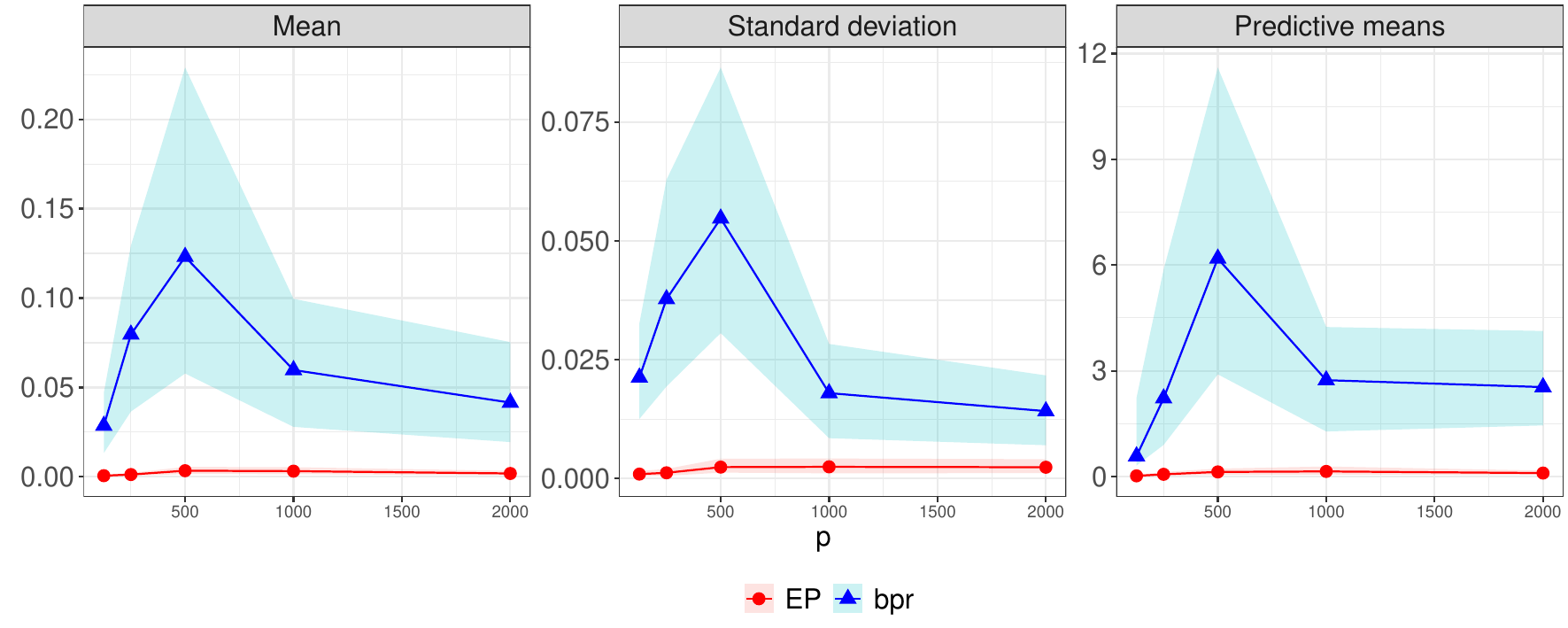}
    \caption{For varying $p$, median absolute difference between the $p$ posterior means, the standard deviations and the $\tilde{n}=100$ predictive probabilities resulting from $10 000$ \textsc{hmc} (\textsc{nuts}) samples and the ones arising from \textsc{ep-eff} and from $10 000$  samples from \texttt{bpr} \citep{dangelo2023efficient} for Poisson regression with $n=500$ and $\nu^{2}=250/p$.
    Shaded areas denote the first and third quartiles.
    }
    \label{fig:2}
\end{figure}

In line with Section \ref{subsec: simPoisson}, we used $10000$ \textsc{hmc} samples obtained from \textsc{nuts}, implementing the Bayesian Poisson model in \texttt{rstan}.
Also in this case, the samples were obtained from $5$ parallel chains of $2000$ samples each, after a burn-in of $1000$.
We thus compared the accuracy of the resulting parameters' posterior means and standard deviations and of the predictive means for the $\tilde{n}=100$ held out units, i.e.\ $\E[\by_{\new}\mid \by]$, resulting from the efficient \ep implementation reported in Algorithms \ref{algo1} and \ref{algo2} (\textsc{ep-eff}) and from the Metropolis-Hastings (\textsc{mh}) scheme developed by \citet{dangelo2023efficient}, implemented in the \texttt{R} package \texttt{bpr}.
This was recently introduced to get posterior samples for Poisson regression models with possibly different choices for the prior distribution, including the Gaussian one.
\texttt{bpr} is particularly efficient for smaller $p$ dimensions, which is indeed the focus of \citet{dangelo2023efficient}.
Conversely, it may become more computationally demanding in higher dimensions due to the intrinsic limitations of \textsc{mh} schemes in high dimensions, as we can also see in Table \ref{table: simPoisson} and Figure \ref{fig:2}.

\begin{table*}[t]
\centering
\caption{\footnotesize{Running time, in seconds, to compute posterior moments and predictive means of the linear predictor for $\tilde{n}=100$ test observations with the \textsc{ep} approximation as in Algorithms \ref{algo1} and \ref{algo2} (\textsc{ep-eff}), with the Metropolis-Hastings sampler from the \texttt{R} package \texttt{bpr} (\texttt{bpr}) and with the \texttt{rstan} implementation of the No-U-Turn \textsc{hmc} sampler (\textsc{nuts}) for Poisson regression with $n=500$ and $\nu^{2}=250/p$.
}}
\label{table: simPoisson}
\begin{tabular}[c]{ll|ccccc}
 \multicolumn{2}{l|}{}  &  \multicolumn{5}{c}{\textit{p}}   \\
\hline
 & \textit{Method} & 125 & 250 & 500 & 1000 & 2000 \\ 
   \hline
Running time &\ \textsc{ep-eff} &\ 0.94 &\ 2.48 &\ 3.14 &\ 5.56 &\ 12.09 \\
(seconds) &\ \texttt{bpr}\ &\ 164.62 &\ 727.21 &\ 4021.31 &\ 25370.04 &\ 184153.96 \\
&\ \textsc{nuts}\ &\ 62.08 &\ 138.64 &\ 580.82 &\ 1082.71 &\ 1672.56 \\
\hline
\end{tabular}
\end{table*}

As shown in Table~\ref{table: simPoisson}, the \textsc{nuts} method is considerably slower, despite the parallel execution of five chains.
As expected, \textsc{ep-eff} significantly minimizes the computational effort required to approximate the posterior moments and predictive means, outperforming the standard sampling methods available.
The computational advantages increase as the problem’s dimensionality grows (i.e., higher values of $p$), resulting in a decrease in running time by more than two orders of magnitude when $p=2000$.
However, when assessing the quality of the approximation of the two posterior moments and predictive means (Figure~\ref{fig:2}), \textsc{ep-eff} consistently provides accurate approximations across different dimensions of $p$, proving to be practically indistinguishable from \textsc{nuts}.
In conjunction with the improved accuracy of \textsc{ep} in all scenarios, this underscores the importance of the proposed efficient implementations for \textsc{ep}.
It renders inference and prediction via \textsc{glm}s computationally feasible in almost every practical scenario, including the challenging high-dimensional and large sample-size settings where previously available implementations are impractical.
Finally, note that for \textsc{ep-eff} we rely on a basic and easily reproducible \texttt{R} implementation.
Optimized computational routines relying on \texttt{C++} coding can be further considered as it is done in \textsc{nuts} and \texttt{bpr}.

\section{Illustrations}
\label{sec: illustration}
\subsection{Probit regression}
\label{subsec: illProbit}
As discussed in \cite{chopin2017leave}, state-of-the-art \textsc{ep} implementations for Bayesian binary regression provide accurate results but are feasible in small-to-moderate $p$ settings, while they tend to become rapidly impractical in large $p$ contexts, such as $p > 1000$.
As a first example to illustrate the improvements provided by the \textsc{ep} algorithm for high-dimensional \textsc{glm}s (Algorithm \ref{algo2}), we consider a large $p$ study to quantify the accuracy and computational costs, expressed as running times, in a real-world application with binary responses.
More precisely, we demonstrate the performance of the \textsc{ep} approximation, derived in Section~\ref{sec: EPprobit}, in a medical application that was previously used in \cite{fasano2022scalable} for comparisons among some state-of-the-art algorithms for approximating posterior and predictive distributions under probit regression.

Following \cite{fasano2022scalable}, we model the presence ($y_{i}=1$) or absence ($y_{i}=0$) of Alzheimer's disease in its early stages as a function of demographic data, genotype, and assay results via an interpretable probit regression.
We further increase flexibility by considering pairwise interactions in the design matrix, $\bX$, thus obtaining $p=9036$ predictors, including the intercept, for $333$ subjects.
The original dataset is available in the \texttt{R} library \texttt{AppliedPredictiveModeling} \citep{craig2011multiplexed}.
The predictors have been standardized to have mean $0$ and standard deviation $0.5$, before being included in the probit models, following a common recommendation in probit regression  \citep[][]{gelman2008weakly,chopin2017leave,fasano2022scalable}.
As for the simulation study, we perform Bayesian inference by employing independent Gaussian priors with mean $0$ and variance $\nu^{2}=25$ for each $\beta_{j}$ $(j=1, \ldots, 9036)$ \citep{gelman2008weakly}.
These priors are updated with the probit likelihood of $n=300$ units, after excluding $33$ subjects to study the behavior of the predictive probabilities in such large~$p$ settings, along with the performance of the overall approximation of the posterior.

\begin{figure}
    \centering
    \includegraphics[width=\linewidth, height=4.5cm]{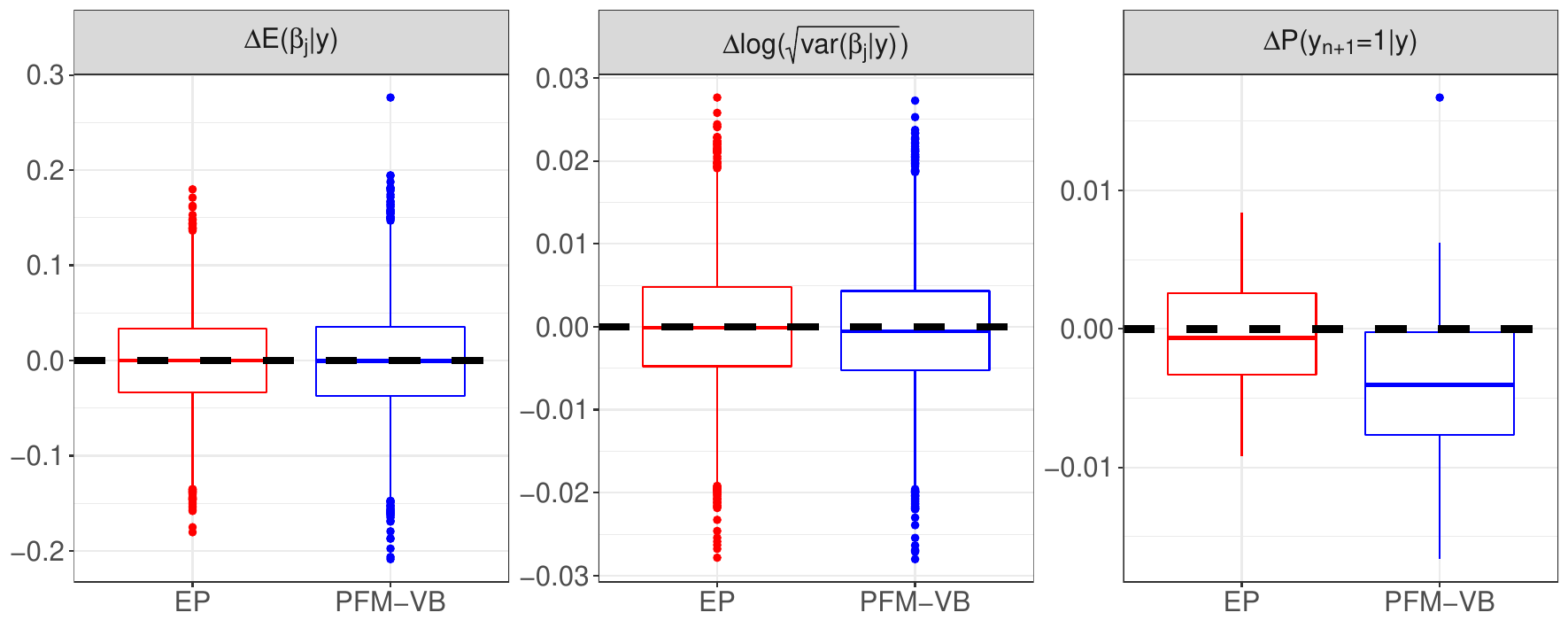}
    \caption{Results for the Alzheimer data under probit regression with $\nu^{2}=25$.
    Boxplot of the differences of the $\E[\beta_{1}\mid\by], \ldots, \E[\beta_{p}\mid\by]$, $\log\big(\sqrt{\text{var}[\beta_{1}\mid\by]}\,\big), \ldots, \log\big(\sqrt{\text{var}[\beta_{p}\mid\by]}\,\big)$, and $\Pr[y_{n+1} =1\mid \by], \ldots, \Pr[y_{n+100} =1\mid \by]$ computed with \textsc{ep} and the \textsc{pfm-vb} solutions, using the inferences obtained via $10 000$ i.i.d.\ samples from the exact \textsc{sun} as benchmark.
    }
    \label{fig: illu probit}
\end{figure}

We compare the accuracy of \textsc{ep}, implemented according to Algorithm~\ref{algo2}, and the \textsc{pfm-vb} approximation \citep{fasano2022scalable} considered in Section \ref{subsec: simProbit} for recovering functionals of interest of the posterior distribution as well as the predictive probabilities for the $33$ held-out units.
The dimensions of the problem --- i.e., large $p$ and moderate $n$ --- make the i.i.d.\ sampler developed by \cite{durante2019conjugate} feasible, and thus a preferable solution over \textsc{mcmc} schemes.
Indeed, since the \textsc{nuts} \textsc{hmc} sampler discussed in Section \ref{subsec: simProbit} takes more than six hours to run in such an example, comparisons are conducted using as a benchmark the results obtained from $10^{4}$ i.i.d.\ samples from \cite{durante2019conjugate}.
As illustrated in Section \ref{subsec: simProbit} and by theoretical results in \cite{fasano2022scalable}, we are focusing on the best scenario for the \textsc{pfm-vb}, i.e., $p\gg n$ since we have already shown in simulation study that \ep outperform \textsc{pfm-vb} in term of accuracy in $n \ge p$ settings.
In such a high-dimensional setting, with $p \gg 1000$, however, routinely-implemented functions, such as \texttt{EPprobit} in the \texttt{EPglm} \texttt{R} package, are not practically usable, as they require more than six hours to run (see also \citet{fasano2022scalable}).
This also highlights the improvements of the proposed efficient \ep implementation derived in Algorithm~\ref{algo2}, which instead requires only a few seconds, see Table \ref{table: illProbit}.

The associate results are illustrated in Figure \ref{fig: illu probit}, where we show the boxplots of the differences between the posterior means $\E[\beta_{1}\mid\by],\ldots,\E[\beta_{p}\mid\by]$, log standard deviations $\log\big(\sqrt{\text{var}[\beta_{1}\mid\by]}\,\big), \ldots, \allowbreak \log\big(\sqrt{\text{var}[\beta_{p}\mid\by]}\,\big)$, and predictive probabilities $\Pr[y_{n+1} =1\mid \by], \ldots, \Pr[y_{n+100} =1\mid \by]$ obtained via the i.i.d.\ sampler and the ones resulting from the two approximate methods.
As expected, both approximate solutions are extremely accurate for all the quantities of interest, which are almost indistinguishable from the ones obtained with the i.i.d.\ sampler.
\ep shows even better accuracy for the posterior means and predictive probabilities.

Finally, Table \ref{table: illProbit} provides insights into the computational time of the different algorithms.
The approximated methods enable the computation of the desired quantities in a fraction of the time required by the i.i.d.\ sampler, which takes approximately $25$ minutes.
Although slightly slower than \textsc{pfm-vb}, which takes only approximately $3$ seconds to produce the results, the proposed implementation of \ep developed in Algorithm \ref{algo2} takes about $12$ seconds to run, paving the way to the use of \ep in high-dimensional Bayesian \textsc{glm}s and showing massive improvements in running times over the available \ep routine for the probit model (\texttt{EPprobit}), which could not run within six hours.

\begin{table*}[t]
\centering
\caption{\footnotesize{Running time, in seconds, to compute posterior moments and predictive probabilities for $\tilde{n}=33$ test observations with the Monte Carlo sampler (\textsc{mc}), with the \texttt{rstan} implementation of the No-U-Turn \textsc{hmc} sampler (\textsc{nuts}), with the \textsc{ep} approximation as in Algorithms \ref{algo1} and \ref{algo2} (\textsc{ep-eff}), with the \textsc{ep} approximation computed via the \texttt{R} function \texttt{EPprobit} (\texttt{EPprobit}), and with the \textsc{pfm-vb} approximation (\textsc{pfm-vb}) for probit regression on the Alzheimer dataset ($n=300$) with $\nu^{2}=25$.
}}
\label{table: illProbit}
\begin{tabular}[c]{l|ccccc}
 &  \multicolumn{3}{c}{Method}   \\
\hline
& \textsc{nuts} & \texttt{EPprobit} & \textsc{mc} & \textsc{ep-eff} & \textsc{pfm-vb}\\
\hline
 Running time (in seconds) & > 21600 & > 21600 & 1541.76 & 12.31 & 3.17\\ 
\hline
\end{tabular}
\end{table*}

\subsection{Poisson regression}
\label{subsec: illPoisson}

To further illustrate the improvements provided by the \textsc{ep} algorithm for high-dimensional \textsc{glm}s (Algorithm \ref{algo2}), we undertake a $p \gg n$ study to quantify the accuracy and computational costs, measured as running times, in a real-world application with integer responses.
Specifically, we showcase the performance of the \textsc{ep} approximation for \textsc{glm}s, derived in Algorithm \ref{algo2} (with the efficient hybrid moment approximation computed in Section \ref{sec: EPhybrid}), in a sports analytics application that was previously used in \cite{dangelo2023efficient} for comparisons among state-of-the-art sampling algorithms under Poisson log-linear regression.
We analyze data of match scores from the Italian Serie A 2020–2021 season, which are publicly available at \url{http://www.football-data.co.uk}.
The number of goals is considered as the variable of interest ($y_{i} \in \mathbb{N}$), while the team, several betting odds (for different betting types and bookmakers), and an indicator of whether the team is playing at home are included as covariates.
We enhance the flexibility in the design matrix $\bX$, by considering pairwise interactions between the teams and the other covariates in the source dataset, thus obtaining $p=1762$ predictors, including the intercept, for $760$ observations.

\begin{figure}[ht]
    \centering
    \includegraphics[width=\linewidth, height=4.5cm]{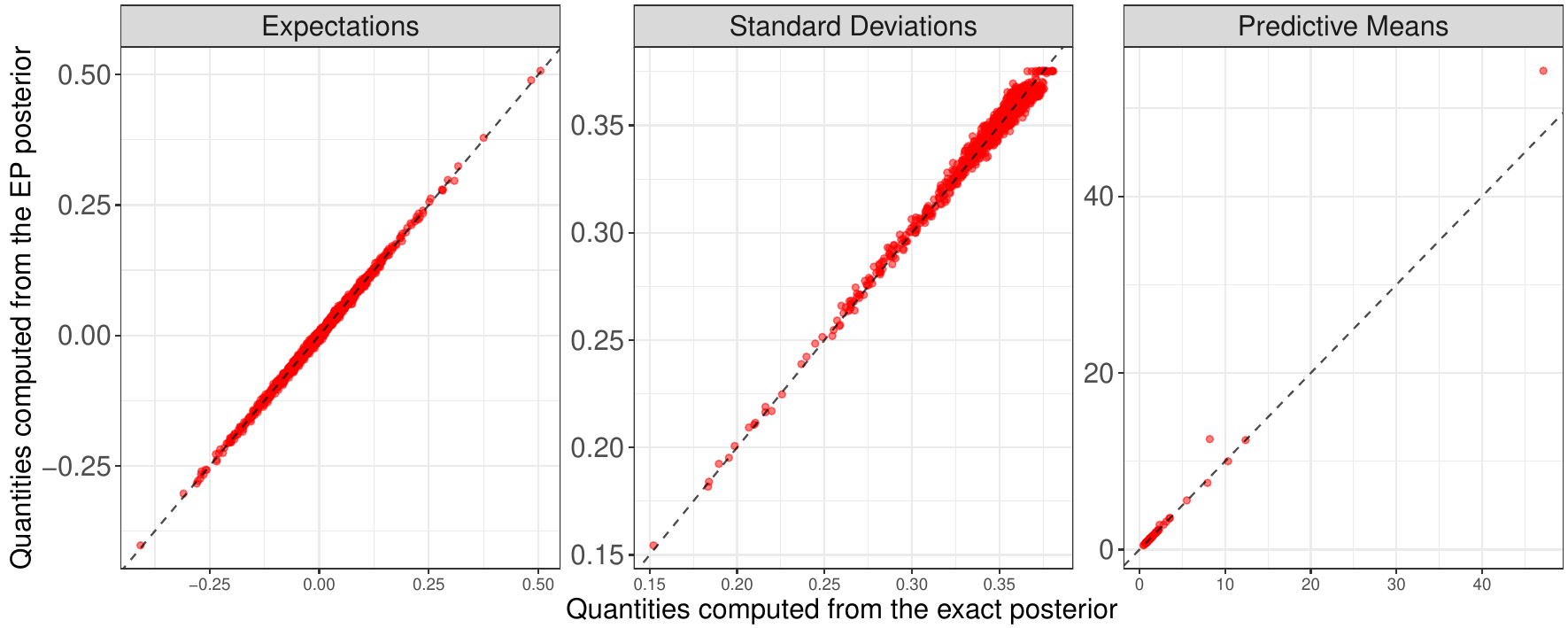}
    \caption{Results for the football data under Poisson regression with $\nu^{2}=250/1762$.
    Scatterplot of posterior expectations and standard deviations of the parameters and predictive means, i.e., $\E[y_{n+1}\mid \by], \ldots, \E[y_{n+60}\mid \by]$, for $60$ held-out units computed via \textsc{nuts} (x-axis) and via the efficient \ep implementation reported in Algorithm \ref{algo2} (y-axis).
    }
    \label{fig: illu poisson}
\end{figure}

As for the simulation studies with count responses, we conduct Bayesian inference by employing independent Gaussian priors with mean $0$ and variance $\nu^{2}=250/p$ for each $\beta_{j}$ $(j=1, \ldots, 1762)$.
These priors are updated with the Poisson likelihood (under logarithm link) of $n=700$ units, after excluding $60$ observations to study the behavior of the predictive means and posterior moments in such $p \gg n$ setting.
In such a high-dimensional scenario, both \texttt{bpr} and \texttt{rstan} incurred insurmountable mixing issues.
To obtain adequate \textsc{mcmc} inference, it was necessary to modify the metric within the momentum term of \textsc{hmc} \citep{betancourt2011geometry}.
In particular, we run \textsc{nuts} via the shell interface \texttt{CmdStan} of \texttt{stan} \citep{gabry2023cmdstanr}, initializing the mass matrix to the approximate posterior variance learned by \textsc{ep}.
Even so, \textsc{nuts} is considerably slower than \textsc{ep}, despite running five shorter chains in parallel as done in the simulation studies.
To produce the results reported in Figure \ref{fig: illu poisson}, the running time of \textsc{nuts} is $\bf{46154.41}$ \textbf{secs} compared to the $\bf{19.56}$ \textbf{secs} of our \textsc{epp-eff} implementation.

Results are depicted in Figure \ref{fig: illu poisson}, where we present the scatter plots of the posterior means $\E[\beta_{1}\mid\by],\ldots,\E[\beta_{p}\mid\by]$, log standard deviations $\log\big(\sqrt{\text{var}[\beta_{1}\mid\by]}\, \big), \ldots, \, \allowbreak \log\big(\sqrt{\text{var}[\beta_{p}\mid\by]}\,\big)$, and predictive means $\E[y_{n+1}\mid \by], \ldots, \E[y_{n+60}\mid \by]$ obtained via the \textsc{nuts} sampler and the ones resultant from the efficient \ep approximation derived in Algorithm \ref{algo2} (with the efficient hybrid moment approximation for Poisson computed in Section \ref{sec: EPhybrid}).
Despite being computationally much faster, the \ep solution is extremely accurate for all the quantities of interest, which are almost identical to the ones obtained with the \textsc{nuts} sampler in most of the cases.

\section{Discussion}\label{sec: discussion}
This contribution develops a general and scalable implementation of \textsc{ep} for the Bayesian \textsc{glm}s under spherical Gaussian prior, making such approximation feasible in scenarios that were not practically possible before \citep{chopin2017leave}.
Such an algorithm provides an approximation of the posterior moments and the marginal likelihood, which may be of practical use for model selection \citep{kass1995bayes}.
Moreover, being the global approximation Gaussian, this also facilitates the computation of \ep predictive mean $\E[y_{\new} \mid \by]$, which is available in closed-form at no additional cost for the probit model (where it coincides with the predictive probability $\Pr[y_{\new}=1 \mid \by]$), and any \textsc{glm} with a logarithmic link (that is the canonical choice for Poisson and Gamma likelihoods).
We further improved computational scalability by considering tailored approximations for the moment-matching step, while preserving accuracy.
As shown on simulated and real data in Sections~\ref{sec: simstudies} and~\ref{sec: illustration}, this decreased cost leads to a massive reduction of the running time compared to currently available implementations of \textsc{ep}, paving the way to applying such a method in high-dimensional settings.
Considering the quality of the approximation empirically observed in the present work, in \citet{chopin2017leave}, and in \citet{anceschi2023bayesian}, this represents a major contribution to Bayesian data regression with responses beyond the continuous case.
In accordance with the robust evidence in the literature, \textsc{ep} seems, in general, the preferable choice among approximate methods.
Finally, these empirical results can also motivate future theoretical research about the goodness of the approximation provided by \textsc{ep}, for which some preliminary findings on the asymptotic behavior have been derived in \citet{dehaene2015bounding} and \citet{dehaene2018expectation}.

\printbibliography

\newpage

\clearpage\pagebreak\newpage
\pagestyle{fancy}
\fancyhf{}
\rhead{\bfseries\thepage}
\lhead{\bfseries SUPPLEMENTARY MATERIALS}

\baselineskip=27pt
\begin{center}
{\LARGE{Supplementary Materials for\\} 
\bf  Scalable expectation propagation \\ for generalized linear models
}
\end{center}

\baselineskip=12pt
\vskip 2mm

\begin{center}
 Niccolò Anceschi$^{1}$ (niccolo.anceschi@duke.edu)\\
    Augusto Fasano$^{2}$ (augusto.fasano@unicatt.it)\\
    Beatrice Franzolini$^{3}$ (beatrice.franzolini@unibocconi.it)\\
    Giovanni Rebaudo$^{4}$ (giovanni.rebaudo@unito.it)
    \vskip 3mm
    $^{1}$Duke University, Durham, USA\\
    $^{2}$Catholic University of the Sacred Heart, Milan, Italy\\
    $^{3}$Bocconi University, Milan, Italy\\
    $^{4}$University of Torino, Turin, Italy
\end{center}

\setcounter{equation}{0}
\setcounter{page}{1}
\setcounter{table}{1}
\setcounter{figure}{0}
\setcounter{section}{0}
\numberwithin{table}{section}
\renewcommand{\theequation}{S.\arabic{equation}}
\renewcommand{\thesubsection}{S.\arabic{section}.\arabic{subsection}}
\renewcommand{\thesection}{S.\arabic{section}}
\renewcommand{\theThm}{S.\arabic{Thm}}
\renewcommand{\theCor}{S.\arabic{Cor}}
\renewcommand{\theProp}{S.\arabic{Prop}}
\renewcommand{\theLem}{S.\arabic{Lem}}
\renewcommand{\thepage}{S.\arabic{page}}
\renewcommand{\thetable}{S.\arabic{table}}
\renewcommand{\thefigure}{S.\arabic{figure}}

\baselineskip=14pt 
\vskip 10mm

\renewcommand{\theequation}{\thesection\arabic{equation}}

\newrefsection
\section{Derivation of the results in Section \ref*{sec: EPglm}}
\label{appendix: MGF}

\subsubsection*{Proof of Lemma \ref*{lemma: hybridMoments} and Proposition \ref*{prop: momentUpdate}}

Adapting the rationale from \citet{zhou2023fast}, the (\textsc{mgf}) $\mathcal{M}_{h_{i}}(\bt)$ of the hybrid distribution $h_{i}(\bbeta)$ reads
\begin{equation*}
\begin{split}
    \mathcal{M}_{h_{i}}(\bt) &= \E_{h_{i}(\bbeta)} \big[ e^{\bbeta^{\intercal} \bt} \big] = \int e^{\bbeta^{\intercal} \bt} \textstyle{\frac{1}{Z_{h_{i}}}} \ell_{i}(\bx_{i}^{\intercal} \bbeta) \phi_{p}(\bbeta - \bxi_{i}, \bOmega_{i}) d\bbeta \\
    &= \textstyle{\frac{1}{Z_{h_{i}}}} e^{\bxi_{i}^{\intercal} \bt + {\textstyle \frac{1}{2}} \bt^{\intercal} \bOmega_{i} \bt} \int \ell_{i}(\eta_{i}) \phi\big(\eta_{i} - \bx_{i}^{\intercal}(\bxi_{i} + \bOmega_{i} \bt), \bx_{i}^{\intercal} \bOmega_{i} \bx_{i} \big) d\eta_{i} \;,
\end{split}
\end{equation*}
for any $\bt \in \R^p$, with $\bOmega_{i} = \bQ_{-i}^{-1}$ and $\bxi_{i} = \bQ_{-i}^{-1} \br_{-i}$.
The last identity follows from $\eta_{i} = \bxi_{i}^{\intercal} \btheta$ and the linear subspace properties of Gaussian measure.
Accordingly, the moments of the hybrid can be obtained as
\begin{equation*}
\begin{split}
    \bmu_{h_{i}} &= \E_{h_{i}(\bbeta)} [ \bbeta ] = \nabla_\bt \mathcal{M}_{h_{i}}(\bzero) \\
    \bSigma_{h_{i}} &= \operatorname{var}_{h_{i}(\bbeta)} [\bbeta] = \nabla_\bt^{2} \mathcal{M}_{h_{i}}(\bzero) - \nabla_\bt \mathcal{M}_{h_{i}}(\bzero) \nabla_\bt^{\intercal} \mathcal{M}_{h_{i}}(\bzero) \; .
\end{split}
\end{equation*}
For ease of notation, let us define $\mathcal{F}_{h_{i}}(\bt) = \int \ell_{i}(\eta_{i}) \phi\big(\eta_{i} - \bx_{i}^{\intercal}(\bxi_{i} + \bOmega_{i} \bt), \bx_{i}^{\intercal} \bOmega_{i} \bx_{i} \big) d\eta_{i}$ and $\mathcal{G}_{h_{i}}(\bt) = e^{\bxi_{i}^{\intercal} \bt + {\textstyle \frac{1}{2}} \bt^{\intercal} \bOmega_{i} \bt}$, so that
\begin{equation*}
\begin{split}
    \nabla_\bt \mathcal{M}_{h_{i}}(\bt) &= \textstyle{\frac{1}{Z_{h_{i}}}} \mathcal{F}_{h_{i}}(\bt) \nabla_\bt \mathcal{G}_{h_{i}}(\bt) + \textstyle{\frac{1}{Z_{h_{i}}}} \mathcal{G}_{h_{i}}(\bt) \nabla_\bt \mathcal{F}_{h_{i}}(\bt) \\
    \nabla_\bt^{2} \mathcal{M}_{h_{i}}(\bt) &= \textstyle{\frac{1}{Z_{h_{i}}}} \mathcal{F}_{h_{i}}(\bt) \nabla_\bt^{2} \mathcal{G}_{h_{i}}(\bt) + \textstyle{\frac{1}{Z_{h_{i}}}} \mathcal{G}_{h_{i}}(\bt) \nabla_\bt^{2} \mathcal{F}_{h_{i}}(\bt) \\
    & \quad + \textstyle{\frac{1}{Z_{h_{i}}}} \nabla_\bt \mathcal{F}_{h_{i}}(\bt) \nabla_\bt^{\intercal} \mathcal{G}_{h_{i}}(\bt) + \textstyle{\frac{1}{Z_{h_{i}}}} \nabla_\bt \mathcal{G}_{h_{i}}(\bt) \nabla_\bt^{\intercal} \mathcal{F}_{h_{i}}(\bt) \; .
\end{split}
\end{equation*}
Cleary, it holds that $ \mathcal{F}_{h_{i}}(\bzero) = Z_{h_{i}} $ and $\mathcal{G}_{h_{i}}(\bzero) = 1 $, while it can be seen that $\nabla_\bt \mathcal{G}_{h_{i}}(\bt) = (\bxi_{i} + \bOmega_{i} \bt)  \mathcal{G}_{h_{i}}(\bt)$ and $\nabla_\bt^{2} \mathcal{G}_{h_{i}}(\bt) = \bOmega_{i} \mathcal{G}_{h_{i}}(\bt) + (\bxi_{i} + \bOmega_{i} \bt) (\bxi_{i} + \bOmega_{i} \bt)^{\intercal} \mathcal{G}_{h_{i}}(\bt) $.
Accordingly, one has
\begin{equation*}
\nabla_\bt \mathcal{G}_{h_{i}}(\bzero) = \bxi_{i} \qquad \qquad \nabla_\bt^{2} \mathcal{G}_{h_{i}}(\bzero) = \bOmega_{i} + \bxi_{i} \bxi_{i}^{\intercal} \; .
\end{equation*}
Conversely, 
\begin{equation*}
\begin{split}
    \nabla_\bt \mathcal{F}_{h_{i}}(\bt) &= \int \ell_{i}(\eta_{i}) \phi\big(\eta_{i} - \bx_{i}^{\intercal}(\bxi_{i} + \bOmega_{i} \bt), \bx_{i}^{\intercal} \bOmega_{i} \bx_{i} \big) \, \\
    & \qquad \Big( \bOmega_{i} \bx_{i} (\bx_{i}^{\intercal} \bOmega_{i} \bx_{i})^{-1} \big(\eta_{i} - \bx_{i}^{\intercal} \bxi_{i} - \bx_{i}^{\intercal} \bOmega_{i} \bt \big) \Big) d\eta_{i} \\
    \nabla_\bt^{2} \mathcal{F}_{h_{i}}(\bt) &= - \int \ell_{i}(\eta_{i}) \phi\big(\eta_{i} - \bx_{i}^{\intercal}(\bxi_{i} + \bOmega_{i} \bt), \bx_{i}^{\intercal} \bOmega_{i} \bx_{i} \big) \, \bOmega_{i} \bx_{i} (\bx_{i}^{\intercal} \bOmega_{i} \bx_{i})^{-1} \bx_{i}^{\intercal} \bOmega_{i} \, d\eta_{i} \\
    &\quad\, + \int \ell_{i}(\eta_{i}) \phi\big(\eta_{i} - \bx_{i}^{\intercal}(\bxi_{i} + \bOmega_{i} \bt), \bx_{i}^{\intercal} \bOmega_{i} \bx_{i} \big) \, \\
    & \qquad \quad \Big( \bOmega_{i} \bx_{i} (\bx_{i}^{\intercal} \bOmega_{i} \bx_{i})^{-1} \big(\eta_{i} - \bx_{i}^{\intercal} \bxi_{i} - \bx_{i}^{\intercal} \bOmega_{i} \bt \big) \Big) \\
    & \qquad \quad \Big( \bOmega_{i} \bx_{i} (\bx_{i}^{\intercal} \bOmega_{i} \bx_{i})^{-1} \big(\eta_{i} - \bx_{i}^{\intercal} \bxi_{i} - \bx_{i}^{\intercal} \bOmega_{i} \bt \big) \Big)^{\intercal} d\eta_{i} \; ,
\end{split} 
\end{equation*}
so that
\begin{equation*}
\begin{split}
    \nabla_\bt \mathcal{F}_{h_{i}}(\bzero) &= Z_{h_{i}} \bOmega_{i} \bx_{i} (\bx_{i}^{\intercal} \bOmega_{i} \bx_{i})^{-1} \big( \E_{h_{i}(\bbeta)} [ \bx_{i}^{\intercal} \bbeta ]  - \bx_{i}^{\intercal} \bxi_{i} \big) \\
    \nabla_\bt^{2} \mathcal{F}_{h_{i}}(\bzero) &= - Z_{h_{i}} \bOmega_{i} \bx_{i} (\bx_{i}^{\intercal} \bOmega_{i} \bx_{i})^{-1} \bx_{i}^{\intercal} \bOmega_{i} + Z_{h_{i}} \bOmega_{i} \bx_{i} (\bx_{i}^{\intercal} \bOmega_{i} \bx_{i})^{-1} \; \cdot \\
    & \quad\; \cdot \Big( \E_{h_{i}(\bbeta)} \big[ (\bx_{i}^{\intercal} \bbeta)^{2} \big] - 2 \E_{h_{i}(\bbeta)} [ \bx_{i}^{\intercal} \bbeta] \bx_{i}^{\intercal} \bxi_{i} + (\bx_{i}^{\intercal} \bxi_{i})^{2} \Big) (\bx_{i}^{\intercal} \bOmega_{i} \bx_{i})^{-1} \bx_{i}^{\intercal} \bOmega_{i} \; .
\end{split}
\end{equation*}
Putting it all together, one gets
\begin{equation*}
\begin{split}
    \nabla_\bt & \mathcal{M}_{h_{i}}(\bzero) = \bxi_{i} + \bOmega_{i} \bx_{i} (\bx_{i}^{\intercal} \bOmega_{i} \bx_{i})^{-1} \big( \E_{h_{i}(\bbeta)} [ \bx_{i}^{\intercal} \bbeta ]  - \bx_{i}^{\intercal} \bxi_{i} \big) \\
    \nabla_\bt^{2} & \mathcal{M}_{h_{i}}(\bzero) = \bxi_{i} \bxi_{i}^{\intercal} + \bOmega_{i} - \bOmega_{i} \bx_{i} (\bx_{i}^{\intercal} \bOmega_{i} \bx_{i})^{-1} \bx_{i}^{\intercal} \bOmega_{i} \\
    & + \bOmega_{i} \bx_{i} (\bx_{i}^{\intercal} \bOmega_{i} \bx_{i})^{-1} \Big( \E_{h_{i}(\bbeta)} \big[ (\bx_{i}^{\intercal} \bbeta)^{2} \big] - 2 \E_{h_{i}(\bbeta)} [ \bx_{i}^{\intercal} \bbeta] \bx_{i}^{\intercal} \bxi_{i} + (\bx_{i}^{\intercal} \bxi_{i})^{2} \Big) (\bx_{i}^{\intercal} \bOmega_{i} \bx_{i})^{-1} \bx_{i}^{\intercal} \bOmega_{i} \\
    & + \big( \bOmega_{i} \bx_{i} \bxi_{i}^{\intercal} + \bxi_{i} \bx_{i}^{\intercal} \bOmega_{i} \big) (\bx_{i}^{\intercal} \bOmega_{i} \bx_{i})^{-1} \big( \E_{h_{i}(\bbeta)} [ \bx_{i}^{\intercal} \bbeta ]  - \bx_{i}^{\intercal} \bxi_{i} \big)
\end{split}
\end{equation*}
and 
\begin{equation*}
\begin{split}
    \nabla_\bt^{2} & \mathcal{M}_{h_{i}}(\bzero) - \nabla_\bt \mathcal{M}_{h_{i}}(\bzero) \nabla_\bt^{\intercal} \mathcal{M}_{h_{i}}(\bzero) = \bOmega_{i} - \bOmega_{i} \bx_{i} (\bx_{i}^{\intercal} \bOmega_{i} \bx_{i})^{-1} \bx_{i}^{\intercal} \bOmega_{i} \\
    & + \bOmega_{i} \bx_{i} (\bx_{i}^{\intercal} \bOmega_{i} \bx_{i})^{-1} \Big( \E_{h_{i}(\bbeta)} \big[ (\bx_{i}^{\intercal} \bbeta)^{2} \big] - \E_{h_{i}(\bbeta)} [ \bx_{i}^{\intercal} \bbeta ]^{2} \Big) (\bx_{i}^{\intercal} \bOmega_{i} \bx_{i})^{-1} \bx_{i}^{\intercal} \bOmega_{i} \; .
\end{split}
\end{equation*}
or equivalently that
\begin{equation*}
\begin{split}
    \bmu_{h_{i}} &= \bxi_{i} +  \bOmega_{i}\bx_{i} (\bx_{i}^{\intercal} \bOmega_{i}\bx_{i})^{-1} \big( \E_{h_{i}(\bbeta)} [\bx_{i}^{\intercal}\bbeta] - \bx_{i}^{\intercal} \bxi_{i} \big)  \\
    \bSigma_{h_{i}} &= \bOmega_{i} + \bOmega_{i}\bx_{i} (\bx_{i}^{\intercal} \bOmega_{i}\bx_{i})^{-1} \big( \Var_{h_{i}(\bbeta)}[\bx_{i}^{\intercal}\bbeta] - \bx_{i}^{\intercal} \bOmega_{i}\bx_{i} \big) (\bx_{i}^{\intercal} \bOmega_{i}\bx_{i})^{-1} \bx_{i}^{\intercal} \bOmega_{i} \; .
 \end{split}
\end{equation*}
Exploiting Woodbury's identity, the natural parameters for the updated global \textsc{ep} approximant read
\begin{equation*}
\begin{split}
    \bQ_{\textsc{ep}}^{\new} &= \big( \Var_{h_{i}(\bbeta)}[\bbeta] \big)^{-1} \\
    &= \bOmega_{i}^{-1} - \bx_{i} \Big( (\bx_{i}^{\intercal} \bOmega_{i}\bx_{i}) \big( \Var_{h_{i}(\bbeta)}[\bx_{i}^{\intercal}\bbeta]  - \bx_{i}^{\intercal} \bOmega_{i}\bx_{i} \big)^{-1} (\bx_{i}^{\intercal} \bOmega_{i}\bx_{i}) \Big)^{-1} \bx_{i}^{\intercal} \\
    & = \bOmega_{i}^{-1} - \bx_{i} \bx_{i}^{\intercal} \big( (\bx_{i}^{\intercal} \bOmega_{i}\bx_{i})^{-1} - \Var_{h_{i}(\bbeta)}[\bx_{i}^{\intercal}\bbeta]^{-1} \big)
\end{split}
\end{equation*}
and 
\begin{equation*}
\begin{split}
    \br_{\textsc{ep}}^{\new} &= \big( \Var_{h_{i}(\bbeta)}[\bbeta] \big)^{-1} \E_{h_{i}(\bbeta)} [\bbeta] \\
    &= \bOmega_{i}^{-1} \bxi_{i} + \bx_{i} \bx_{i}^{\intercal} \big( \Var_{h_{i}(\bbeta)}[\bx_{i}^{\intercal} \bbeta]^{-1} \E_{h_{i}(\bbeta)} [\bx_{i}^{\intercal} \bbeta] -  (\bx_{i}^{\intercal} \bOmega_{i}\bx_{i})^{-1} \bx_{i}^{\intercal} \bxi_{i} \big) \; .
\end{split}
\end{equation*}
Together with the definition of the cavity parameters as $\bOmega_{i} = \bQ_{-i}^{-1}$ and $\bxi_{i} = \bQ_{-i}^{-1} \br_{-i}$, this gives the one-dimensional updates for the $i$-th site, reported in Proposition \ref*{prop: momentUpdate}.

\subsubsection*{Proof of Proposition \ref*{prop: updateZi}}
Specifying equation (\ref*{eq: updateZi}),
\begin{align}
\label{eq: updateZiProbitComputations}
    \log Z_{i}^{\new} &= \dfrac{1}{2} \br_{\ep}^{\new\ \intercal} \bOmega_{\ep}^{\new} \br_{\ep}^{\new} +\dfrac{p}{2}\log(2\pi) - \dfrac{1}{2}\log |\bQ_{\ep}^{\new}|\\ \nonumber
    &\quad \quad -\dfrac{1}{2} \br_{-i}^{\intercal} \bOmega_{i} \br_{-i} -\dfrac{p}{2}\log(2\pi) + \dfrac{1}{2}\log |\bQ_{-i}| - \log Z_{h_{i}} \\ 
    &=\dfrac{1}{2} \left(\br_{\ep}^{\new\ \intercal} \bOmega_{\ep}^{\new} \br_{\ep}^{\new} - \br_{-i}^{\intercal} \bOmega_{i} \br_{-i}\right) 
    -\dfrac{1}{2} \left(\log |\bQ_{\ep}^{\new}| - \log |\bQ_{-i}| \right) - \log Z_{h_{i}}. \nonumber
\end{align}
Now, for the first parenthesis, exploiting Woodbury's identity, one gets
\begin{equation*}
\begin{split}
    &= \br_{\ep}^{\new\ \intercal} \bOmega_{\ep}^{\new} \br_{\ep}^{\new} - \br_{-i}^{\intercal} \bOmega_{i} \br_{-i} \\ 
    &= (\br_{-i}+m_{i}^{\new}\bx_{i})^{\intercal}(\bQ_{-i}+k_{i}^{\new}\bx_{i}\bx_{i}^{\intercal})^{-1}(\br_{-i}+m_{i}^{\new}\bx_{i}) - \br_{-i}^{\intercal} \bOmega_{i} \br_{-i}\\
    &=(\br_{-i}+m_{i}^{\new}\bx_{i})^{\intercal}\left[\bOmega_{i}-\dfrac{k_{i}^{\new}}{1+k_{i}^{\new} \bx_{i}^{\intercal}\bOmega_{i}\bx_{i}} \bOmega_{i}\bx_{i} \bx_{i}^{\intercal}\bOmega_{i}\right](\br_{-i}+m_{i}^{\new}\bx_{i})\\&\qquad \qquad \qquad \qquad \qquad \qquad \qquad \qquad \qquad \qquad \qquad \qquad - \br_{-i}^{\intercal} \bOmega_{i} \br_{-i}\\
    &=\br_{-i}^{\intercal} \bOmega_{i} \br_{-i} + 2 m_{i}^{\new}\br_{-i}^{\intercal} \bOmega_{i} \bx_{i} + (m_{i}^{\new})^{2}\bx_{i}^{\intercal}\bOmega_{i} \bx_{i}\\
    &\quad \quad -\dfrac{k_{i}^{\new}}{1+k_{i}^{\new} \bx_{i}^{\intercal}\bOmega_{i}\bx_{i}}\Big[(\br_{-i}^{\intercal}\bOmega_{i}\bx_{i} )^{2} +2m_{i}^{\new} (\br_{-i}^{\intercal} \bOmega_{i} \bx_{i}) (\bx_{i}^{\intercal}\bOmega_{i}\bx_{i})\\
    &\qquad \qquad \qquad \qquad \qquad \qquad +(m_{i}\bx_{i}^{\intercal}\bOmega_{i}\bx_{i})^{2}\Big] - \br_{-i}^{\intercal} \bOmega_{i} \br_{-i}\\
    &=2m_{i}^{\new} \br_{-i}^{\intercal} \bOmega_{i} \bx_{i}\left[1-\dfrac{k_{i}^{\new} \bx_{i}^{\intercal}\bOmega_{i} \bx_{i}}{1+k_{i}^{\new}\bx_{i}^{\intercal}\bOmega_{i} \bx_{i}} \right] - \dfrac{k_{i}^{\new} } {1+k_{i}^{\new}\bx_{i}^{\intercal}\bOmega_{i} \bx_{i}} (\br_{-i}^{\intercal}\bOmega_{i}\bx_{i} )^{2}\\
    &\quad \quad + (m_{i}^{\new})^{2} \bx_{i}^{\intercal}\bOmega_{i} \bx_{i} \left[1-\dfrac{k_{i}^{\new} \bx_{i}^{\intercal}\bOmega_{i} \bx_{i}}{1+k_{i}^{\new}\bx_{i}^{\intercal}\bOmega_{i} \bx_{i}} \right]\\
    &=\dfrac{1}{1+k_{i}^{\new}\bx_{i}^{\intercal}\bOmega_{i} \bx_{i}}\left[2m_{i}^{\new} \bx_{i}^{\intercal}\bOmega_{i} \br_{-i}+ (m_{i}^{\new})^{2} \bx_{i}^{\intercal}\bOmega_{i} \bx_{i}  - k_{i}^{\new} (\br_{-i}^{\intercal}\bOmega_{i}\bx_{i} )^{2}\right]
\end{split}
\end{equation*}
As for the second parenthesis, exploiting standard properties of determinants of matrices \citep{petersen2008matrix} it holds
\begin{equation*}
\begin{split}
    \log |\bQ_{\ep}^{\new}| - \log |\bQ_{-i}|&=\log |\bQ_{-i} + k_{i}^{\new}\bx_{i} \bx_{i}^{\intercal}| - \log |\bQ_{-i}|\\
    &=\log |\bQ_{-i}[\bI_{p}+k_{i}^{\new}\bOmega_{i} \bx_{i} \bx_{i}^{\intercal}]|\\
    &=\log |\bQ_{-i}| + \log|\bI_{p}+k_{i}^{\new}\bOmega_{i} \bx_{i} \bx_{i}^{\intercal}| - \log |\bQ_{-i}|\\
    &=\log(1+k^{\new} \bx_{i}^{\intercal} \bOmega_{i} \bx_{i}).
\end{split}
\end{equation*}
Substituting the above derivations in \eqref{eq: updateZiProbitComputations}, one finally obtains
\begin{equation*}
\begin{split}
    \log Z_{i}^{\new} = \dfrac{1}{2}\bigg[
    &\dfrac{1}{1+k_{i}^{\new}\bx_{i}^{\intercal} \bOmega_{i}\bx_{i}}\left(2m_{i}^{\new} \br_{-i}^{\intercal} \bOmega_{i} \bx_{i}+(m_{i}^{\new})^{2} \bx_{i}^{\intercal} \bOmega_{i}\bx_{i} -k_{i}^{\new} (\br_{-i}^{\intercal} \bOmega_{i} \bx_{i})^{2} \right)\\
    &-\log (1+k_{i}^{\new} \bx_{i}^{\intercal} \bOmega_{i} \bx_{i}) \bigg]
    -\log Z_{h_{i}},
\end{split}
\end{equation*}
as claimed.
\qed

\subsubsection*{Proof of Proposition \ref*{prop: updatesLargeP}}

By Woodbury's identity, one gets
\begin{equation*}
	\begin{split}
	\bw_{i} &= \bQ_{-i}^{-1}\bx_{i} = (\bQ-\bQ_{i})^{-1} \bx_{i}= \bQ^{-1}\bx_{i} + (1-k_{i}\bx_{i}^{\intercal}\bQ^{-1}\bx_{i})^{-1} k_{i} (\bQ^{-1}\bx_{i})(\bQ^{-1}\bx_{i})^{\intercal}\bx_{i}\\
	&=\bv_{i}+ k_{i} (1-k_{i}\bx_{i}^{\intercal}\bv_{i})^{-1} \bv_{i}\bv_{i}^{\intercal} \bx_{i}
	= \big[1+ (1-k_{i}\bx_{i}^{\intercal}\bv_{i})^{-1} (k_{i}\bx_{i}^{\intercal} \bv_{i})\big] \bv_{i}
	= d_{i} \bv_{i},
	\end{split}
\end{equation*}
where $d_{i}=(1-k_{i}\bx_{i}^{\intercal}\bv_{i})^{-1}$.

When site $i$ is updated, $\bQ_{i}$ changes and consequently also $\bOmega_{\ep}$ does, assuming a new value $\bOmega_{\ep}^{\new}=(\bQ_{-i}+\bQ_{i}^{\new})^{-1}$.
Thus also the $\bv_{j}$'s, $j=1,\ldots,n$, should be updated.
Exploiting the form of $\bOmega_{\ep}^{\new}$, together with Woodbury's identity, one gets 
\begin{equation*}
\begin{split}
\bv_{j}^{\new} &= \bOmega_{\ep}^{\new} \bx_{j} = (\bQ_{\ep} - \bQ_{i}+\bQ_{i}^{\new})^{-1}\bx_{j}
= [\bQ_{\ep} + (k_{i}^{\new} - k_{i}) \bx_{i} \bx_{i}^{\intercal}]^{-1}\bx_{j}\\
&= [\bOmega_{\ep} - (k_{i}^{\new} - k_{i}) [1+(k_{i}^{\new} - k_{i})\bx_{i}^{\intercal}\bOmega_{\ep}\bx_{i} ]^{-1} \bOmega_{\ep}\bx_{i}\bx_{i}^{\intercal}\bOmega_{\ep} ]\bx_{j}\\
&= \bOmega_{\ep}\bx_{j}-[(k_{i}^{\new} - k_{i})^{-1} +\bx_{i}^{\intercal}\bv_{i} ]^{-1} \bv_{i}\bx_{i}^{\intercal}\bv_{j}
= \bv_{j} - c_{i} (\bx_{i}^{\intercal}\bv_{j}) \bv_{i},
\end{split}
\end{equation*}
where $c_{i}=(k_{i}^{\new} - k_{i})/ (1 + (k_{i}^{\new} - k_{i})\bx_{i}^{\intercal}\bv_{i})$.\qed

\section{Derivation of the results in Section \ref*{sec: EPhybrid}}
\label{appendix: comp}

\subsubsection*{Proof of equation (\ref*{eq: moments from diff log Z})}

Recalling that here $\eta_{i} = \bx_{i}^{\intercal} \bbeta$ and exploring differentiation under integral sign, one has
\begin{equation*}
\begin{split}
    \frac{\partial}{\partial \lambda_{i}} \log Z_{h_{i}} &= \frac{1}{Z_{h_{i}}} \frac{\partial Z_{h_{i}}}{\partial \lambda_{i}} = \frac{1}{Z_{h_{i}}} \int \ell_{i}(\eta_{i}) \frac{\partial}{\partial \lambda_{i}} \phi \big(\eta_{i} - \lambda_{i}, \, \rho_{i}^{2} \big)d\eta_{i} \\
    &= \frac{1}{Z_{h_{i}}} \int \ell_{i}(\eta_{i}) \phi \big(\eta_{i} - \lambda_{i}, \, \rho_{i}^{2} \big) \frac{\partial}{\partial \lambda_{i}} \Big( - \frac{(\eta_{i}-\lambda_{i})^{2}}{2 \rho_{i}^{2}} \Big) d\eta_{i} \\
    &= \frac{1}{Z_{h_{i}}} \int \ell_{i}(\eta_{i}) \phi \big(\eta_{i} - \lambda_{i}, \, \rho_{i}^{2} \big)  \frac{(\eta_{i}-\lambda_{i})}{\rho_{i}^{2}} d\eta_{i} = \frac{1}{\rho_{i}^{2}} \big(\E_{h_{i}(\bbeta)} [\eta_{i}]-\lambda_{i} \big) \; .
\end{split}
\end{equation*}
A simple rearrangement gives $\E_{h_{i}(\bbeta)} [\eta_{i}] = \lambda_{i} + \rho_{i}^{2} \frac{\partial}{\partial \lambda_{i}} \log Z_{h_{i}}$.
Similarly, it holds that
\begin{equation*}
\begin{split}
    \frac{\partial}{\partial \rho_{i}^{2}} \log Z_{h_{i}} &= \frac{1}{Z_{h_{i}}} \frac{\partial Z_{h_{i}}}{\partial \rho_{i}^{2}} = \frac{1}{Z_{h_{i}}} \int \ell_{i}(\eta_{i}) \frac{\partial}{\partial \rho_{i}^{2}} \phi \big(\eta_{i} - \lambda_{i}, \, \rho_{i}^{2} \big)d\eta_{i} \\
    &= \frac{1}{Z_{h_{i}}} \int \ell_{i}(\eta_{i}) \phi \big(\eta_{i} - \lambda_{i}, \, \rho_{i}^{2} \big) \Big( - \frac{1}{2 \rho_{i}^{2}} - \frac{\partial}{\partial \rho_{i}^{2}} \frac{(\eta_{i}-\lambda_{i})^{2}}{2 \rho_{i}^{2}} \Big) d\eta_{i} \\
    &= \frac{1}{Z_{h_{i}}} \int \ell_{i}(\eta_{i}) \phi \big(\eta_{i} - \lambda_{i}, \, \rho_{i}^{2} \big) \frac{1}{2 \rho_{i}^{2}} \Big(  \frac{(\eta_{i}-\lambda_{i})^{2}}{2 \rho_{i}^{2}} - 1 \Big) d\eta_{i} \\
    &= \frac{1}{2 \rho_{i}^{2}} \Big( \frac{1}{\rho_{i}^{2}} \, \E_{h_{i}(\bbeta)} [\eta_{i}^{2}] - 2 \, \frac{1}{\rho_{i}^{2}} \, \lambda_{i} \, \E_{h_{i}(\bbeta)} [\eta_{i}] + \frac{1}{\rho_{i}^{2}} \, \lambda_{i}^{2} \Big) - \frac{1}{2 \rho_{i}^{2}} \\
    &= \frac{1}{2 \rho_{i}^{2}} \Big( \frac{1}{\rho_{i}^{2}} \, \operatorname{var}_{h_{i}(\bbeta)} [\eta_{i}] + \frac{1}{\rho_{i}^{2}} \, \big( \E_{h_{i}(\bbeta)} [\eta_{i}] - \lambda_{i} \big)^{2} \Big) - \frac{1}{2 \rho_{i}^{2}} \\
    &= \frac{1}{2 \rho_{i}^{2}} \Big( \frac{1}{\rho_{i}^{2}} \, \operatorname{var}_{h_{i}(\bbeta)} [\eta_{i}] + \rho_{i}^{2} \big( \frac{\partial}{\partial \lambda_{i}} \log Z_{h_{i}} \big)^{2} \Big) - \frac{1}{2 \rho_{i}^{2}} \; .
\end{split}
\end{equation*}
Trivial rearrangements give $\operatorname{var}_{h_{i}(\bbeta)} [\eta_{i}] = \rho_{i}^{2} + \rho_{i}^{2} \Big( 2 \frac{\partial}{\partial \rho_{i}^{2}} \log Z_{h_{i}} -  \big( \frac{\partial}{\partial \lambda_{i}} \log Z_{h_{i}} \big)^{2} \Big) \rho_{i}^{2}$.
We note that the expression for the variance could have been equivalently derived by studying the second derivative of $\log Z_{h_{i}}$ with respect to $\lambda_{i}^{2}$
\begin{equation*}
    \frac{\partial^{2}}{\partial \lambda_{i}^{2}} \log Z_{h_{i}} =
    \frac{\partial }{\partial \lambda_{i}} \Big( \frac{\partial }{\partial \lambda_{i}} \log  Z_{h_{i}} \Big) = 
    \frac{1}{Z_{h_{i}}} \frac{\partial^{2} Z_{h_{i}}}{\partial \lambda_{i}^{2}} - \Big( \frac{1}{Z_{h_{i}}} \frac{\partial Z_{h_{i}}}{\partial \lambda_{i}} \Big)^{2} \; .
\end{equation*}
Working out the first term
\begin{equation*}
\begin{split}
    \frac{1}{Z_{h_{i}}} \frac{\partial^{2} Z_{h_{i}}}{\partial \lambda_{i}^{2}} &= \frac{1}{Z_{h_{i}}} \int \ell_{i}(\eta_{i}) \frac{\partial^{2}}{\partial \lambda_{i}^{2}} \phi \big(\eta_{i} - \lambda_{i}, \, \rho_{i}^{2} \big)d\eta_{i} \\
    & = \frac{1}{Z_{h_{i}}} \int \ell_{i}(\eta_{i}) \frac{\partial}{\partial \lambda_{i}} \Big( \phi \big(\eta_{i} - \lambda_{i}, \, \rho_{i}^{2} \big) \frac{(\eta_{i}-\lambda_{i})}{\rho_{i}^{2}} \Big) d\eta_{i} \\
    & = \frac{1}{Z_{h_{i}}} \int \ell_{i}(\eta_{i}) \phi \big(\eta_{i} - \lambda_{i}, \, \rho_{i}^{2} \big) \frac{1}{\rho_{i}^{2}} \Big( \frac{(\eta_{i}-\lambda_{i})^{2}}{\rho_{i}^{2}} - 1 \Big) d\eta_{i} \\
    &= \frac{1}{\rho_{i}^{2}} \Big( \frac{1}{\rho_{i}^{2}} \, \E_{h_{i}(\bbeta)} [\eta_{i}^{2}] - 2 \, \frac{1}{\rho_{i}^{2}} \, \lambda_{i} \, \E_{h_{i}(\bbeta)} [\eta_{i}] + \frac{1}{\rho_{i}^{2}} \, \lambda_{i}^{2} \Big) - \frac{1}{\rho_{i}^{2}} \; ,
\end{split}
\end{equation*}
one has that $\frac{\partial}{\partial \rho_{i}^{2}} \log Z_{h_{i}} = \frac{1}{2} \frac{\partial^{2}}{\partial \lambda_{i}^{2}} \log Z_{h_{i}}$.
Nonetheless, we note that this last equality holds true in the exact case, but it typically breaks down for the approximations considered in Section~\ref*{sec: EPhybrid}.
In such cases, dedicated analysis could be performed to assess empirically which of the two alternatives provides a better approximation of the exact hybrid moments.

\subsubsection*{Proof of Proposition \ref*{prop: hybrid probit}}
By definition, the hybrid distribution $h_{i}(\bbeta)$ is proportional to 
\begin{equation*}
	p(y_{i}\mid \bbeta) q_{-i}(\bbeta) =
	\Phi((2y_{i}-1)\bx_{i}^{\intercal}\bbeta)\phi_{p}(\bbeta-\bQ_{-i}^{-1}\br_{-i},\bQ_{-i}^{-1}).
\end{equation*}
Recall that $\bxi_{i} = \bQ_{-i}^{-1}\br_{-i}$ and $\bOmega_{i} = \bQ_{-i}^{-1}$, and define
\begin{equation*}
\begin{split}
    \bomega_{i} &=\left[\text{diag}\left(\bOmega_{i}\right)\right]^{1/2} 
    \hspace*{50pt}
    \bar{\bOmega}_{i} = \bomega_{i}^{-1} \bOmega_{i} \bomega_{i}^{-1} \\
    \balpha_{i} &= (2y_{i}-1)\bomega_{i}\bx_{i}
    \hspace*{55pt}
    \tau_{i} = (2y_{i}-1)(1+\bx_{i}^{\intercal}\bOmega_{i}\bx_{i})^{-1/2}\bx_{i}^{\intercal}\bxi_{i} \; .
\end{split}
\end{equation*}
After some trivial algebraic manipulations, one gets that
\begin{equation*}
    p(y_{i}\mid \bbeta) q_{-i}(\bbeta) = \Phi\left(\tau_{i}(1+\balpha_{i}^{\intercal}\bar{\bOmega}_{i}\balpha_{i})^{1/2} + \balpha_{i}^{\intercal} \bomega_{i}^{-1}(\bbeta-\bxi_{i}) \right) \phi_{p}\left( \bbeta-\bxi_{i},\bOmega_{i} \right),
\end{equation*}
which is the kernel of a multivariate extended skew-normal distribution $\textsc{sn}_{p}(\bxi_{i},\bOmega_{i},\balpha_{i},\tau_{i})$ (see Equation (5.61) in \cite{azzalini2014skew}), with normalizing constant $Z_{h_{i}}=\Phi(\tau_{i})$.
Furthermore, exploiting formulae (5.71) and (5.72) in \cite{azzalini2014skew}, the first two moments of $h_{i}(\bbeta)$ are readily available:
\begin{equation*}
\begin{split}
	\bmu_{h_{i}} &= \E_{h_{i}(\bbeta)} [\bbeta] = \bxi_{i}+\zeta_{1}(\tau_{i}) s_{i}\bOmega_{i}\bx_{i},\\
	\bSigma_{h_{i}} &= \Var_{h_{i}(\bbeta)} [\bbeta] = \bOmega_{i} + \zeta_2(\tau_{i}) s_{i}^{2}(\bOmega_{i}\bx_{i})(\bOmega_{i}\bx_{i})^{\intercal},
\end{split}
\end{equation*}
This allows for an alternative derivation of the \textsc{ep} updates for the $i$-th site one-dimensional parameters, without having to resort to the \textsc{mgf} of the hybrid distribution.
Appealing once more to Woodbury's identity, one has
\begin{equation*}
	\begin{split}
	\bQ_{i}^{\new} & = \bSigma_{h_{i}}^{-1} - \bQ_{-i} \\
	&=\bOmega_{i}^{-1} - \zeta_2(\tau_{i}) s_{i}^{2} (1+\zeta_2(\tau_{i}) s_{i}^{2}\bx_{i}^{\intercal}\bOmega_{i}\bOmega_{i}^{-1}\bOmega_{i}\bx_{i})^{-1}\bOmega_{i}^{-1}\bOmega_{i} \bx_{i} \bx_{i}^{\intercal} \bOmega_{i}\bOmega_{i}^{-1} -\bQ_{-i}\\
	& = -\zeta_2(\tau_{i}) s_{i}^{2} (1+\zeta_2(\tau_{i}) s_{i}^{2}\bx_{i}^{\intercal}\bOmega_{i}\bx_{i})^{-1}\bx_{i} \bx_{i}^{\intercal}
	= -(\zeta_2(\tau_{i})^{-1} s_{i}^{-2} +\bx_{i}^{\intercal}\bOmega_{i}\bx_{i})^{-1} \bx_{i} \bx_{i}^{\intercal}\\
	& = -\dfrac{\zeta_2(\tau_{i})}{1 + \bx_{i}^{\intercal}\bOmega_{i}\bx_{i} + \zeta_2(\tau_{i})\bx_{i}^{\intercal}\bOmega_{i}\bx_{i}} \bx_{i} \bx_{i}^{\intercal}
	= k_{i}^{\new} \bx_{i} \bx_{i}^{\intercal},
	\end{split}
\end{equation*}
with $k_{i}^{\new} = -\zeta_2(\tau_{i})/\left(1 + \bx_{i}^{\intercal}\bOmega_{i}\bx_{i} + \zeta_2(\tau_{i})\bx_{i}^{\intercal}\bOmega_{i}\bx_{i}\right)$.
Then, exploiting the previous result, one gets
\begin{equation*}
	\begin{split}
	\br_{i}^{\new} &= \bSigma_{h_{i}}^{-1} \bmu_{h_{i}} - \br_{-i} \\
        &= \bQ_{-i}\bmu_{h_{i}} + \bQ_{i}^{\new}\bmu_{h_{i}} - \br_{-i}= \bQ_{-i} \bQ_{-i}^{-1} \br_{-i} + \zeta_{1}(\tau_{i}) s_{i} \bQ_{-i}\bOmega_{i}\bx_{i} + \bQ_{i}^{\new}\bmu_{h_{i}} - \br_{-i}\\
	&=  \zeta_{1}(\tau_{i}) s_{i} \bx_{i} + \bQ_{i}^{\new}\bmu_{h_{i}}
	= \zeta_{1}(\tau_{i}) s_{i} \bx_{i} + k_{i}^{\new} \bx_{i} \bx_{i}^{\intercal} \bOmega_{i} \br_{-i} + k_{i}^{\new} \zeta_{1}(\tau_{i}) s_{i} \bx_{i}\bx_{i}^{\intercal} \bOmega_{i}\bx_{i}\\
	&= [\zeta_{1}(\tau_{i}) s_{i} + k_{i}^{\new}   (\bOmega_{i}\bx_{i})^{\intercal} \br_{-i} + k_{i}^{\new} \zeta_{1}(\tau_{i}) s_{i} \bx_{i}^{\intercal} \bOmega_{i}\bx_{i}] \bx_{i}= m_{i}^{\new} \bx_{i},
	\end{split}
\end{equation*}
where $m_{i}^{\new} = \zeta_{1}(\tau_{i}) s_{i} + k_{i}^{\new}   (\bOmega_{i}\bx_{i})^{\intercal} \br_{-i} + k_{i}^{\new} \zeta_{1}(\tau_{i}) s_{i} \bx_{i}^{\intercal} \bOmega_{i}\bx_{i}$.
\qed

\subsubsection*{Approximation of the log-normal Laplace transform from \citet{asmussen2014laplace}}

In this section, we work out the details of the approximation of the Laplace transform for log-normal random variable $\mathcal{L}og\mathcal{N}orm(0, \rho^{2})$, developed in \citet{asmussen2014laplace}.
The latter is given by the expectation 
\begin{equation*}
    \mathcal{L}(s) = \E \big[ e^{- s e^{X_{0}}}\big] = \int_{-\infty}^\infty \frac{1}{\sqrt{2 \pi \rho^{2}} }\exp{\Big(-s e^{x_{0}} -\frac{x_{0}^{2}}{2 \rho^{2}}\Big)} dx_{0}
\end{equation*}
where $X_{0} \sim \mathcal{N}(0, \rho^{2})$, which is defined only for $s>0$.
Extensions to expectations of the form $\E \big[ e^{- s e^{X}}\big]$, for $X \sim \mathcal{N}(\varphi, \rho^{2})$, can be then trivially obtained as $\mathcal{L}(s e^{\varphi})$.
The authors essentially suggest applying the standard Laplace method to the integral corresponding to the required expectation.
In general, this gives
\begin{equation*}
    \int_a^b \frac{1}{\sqrt{2 \pi}} e^{- \gamma f(x_{0})} dx_{0} = 
    \frac{e^{-\gamma f(\hat{x}_{0})}}{\sqrt{\gamma f''(\hat{x}_{0})}} 
    \big( 1 + \mathcal{O}(\gamma^{-1})\big) \qquad \text{as} \qquad \gamma \rightarrow \infty \; ,
\end{equation*}
provided that $f(x_{0})$ has a unique global minimum at  $\hat{x}_{0} \in (a,b)$.
Nonetheless, this approximation can be very accurate not only in asymptotic settings but also for relatively small values of $\gamma$ \citep{asmussen2014laplace}.
Although in our context we lack an explicit parameter $\gamma$, 
an analogous role can be assigned to the second derivative of $f(\cdot)$ at its minimum.
In the case of $\mathcal{L}(s)$, $\gamma f(x_{0})$ gets replaced by $g(x_{0} ; s,  \rho^{2}) = s e^{x_{0}} + \frac{x_{0}^{2}}{2 \rho^{2}}$, whose derivatives are trivially obtained as $g'(x_{0} ; s,  \rho^{2}) = s e^{x_{0}} + \frac{x_{0}}{\rho^{2}}$ and $g''(x_{0} ; s,  \rho^{2}) = s e^{x_{0}} + \frac{1}{\rho^{2}}$.
Note that $g(x_{0} ; s,  \rho^{2}) $ is convex in $x_{0} $ if $s>0$.
Hence, its global minimum $\hat{x}_{0}$ is obtained by setting to zero the first derivative, which gives $-\hat{x}_{0} e^{-\hat{x}_{0}} = s \rho^{2}$.
Recalling that the \textit{Lambert W function} is defined as the solution of $\mathcal{W}(x) e^{\mathcal{W}(x)} = x$, one gets that $\hat{x}_{0} = - \mathcal{W}(s \rho^{2})$.
Accordingly
\begin{equation*}
\begin{split}
    g(\hat{x}_{0} ; s,  \rho^{2}) &= \frac{1}{\rho^{2}} \mathcal{W}(s \rho^{2}) + \frac{1}{2 \rho^{2}} \mathcal{W}(s \rho^{2})^{2} \\
    g''(\hat{x}_{0} ; s,  \rho^{2}) &= \frac{1 + \mathcal{W}(s \rho^{2})}{\rho^{2}} \; .
\end{split}
\end{equation*}
Putting it all together, we Laplace method applied to $\mathcal{L}(s)$ gives the approximation
\begin{equation*}
    \widetilde{\mathcal{L}}(s) = \frac{1}{\sqrt{1 + \mathcal{W}(s \rho^{2})}} \exp{\Big( - \frac{1}{\rho^{2}} \mathcal{W}(s \rho^{2}) - \frac{1}{2 \rho^{2}} \mathcal{W}(s \rho^{2})^{2} \Big)} \; .
\end{equation*}
We further highlight that $\frac{\partial}{\partial x} \mathcal{W}(x) = \frac{\mathcal{W}(x)}{1+\mathcal{W}(x)} \frac{1}{x}$, which is going to be useful in the next two Sections for the computations of the derivatives of $\widetilde{\mathcal{L}}(s)$.

\subsubsection*{Adaptation to approximate moments matching in Poisson regression}

Exploiting the results presented in the previous appendix, it is easy to see that the intractable integral in the normalizing constant of \textsc{ep} hybrids for Poisson regression in (\ref*{eq: norm const poisson}) coincides with $\mathcal{L}(s_{i})$ for $s_{i} = \exp{(\lambda_{i} + \rho_{i}^{2} y_{i})}$.
Accordingly, we approximate it with $\widetilde{\mathcal{L}}(s_{i})$.
As a consequence, one has that 
\begin{equation*}
\begin{split}
   \frac{\partial}{\partial \lambda_{i}} \log \widetilde{Z}_{h_{i}} &= \frac{\partial}{\partial \lambda_{i}} \Big( -\log(y_{i}!) + \lambda_{i} y_{i} + \frac{1}{2} \rho_{i}^{2} y_{i}^{2} + \log \widetilde{\mathcal{L}}(s_{i}) \Big) \\
   &= \frac{\partial}{\partial \lambda_{i}} \Big( \lambda_{i} y_{i} + \frac{1}{2} \rho_{i}^{2} y_{i}^{2} -\dfrac{1}{\rho_{i}^{2}}\mathcal{W}( s_{i} \rho_{i}^{2}) -\dfrac{1}{2\rho_{i}^{2}}\mathcal{W}( s_{i} \rho_{i}^{2}) ^{2}-\dfrac{1}{2}\log \big( 1+ \mathcal{W} ( s_{i} \rho_{i}^{2}) \big) \Big) \\
   &= y_{i} - \Big( \dfrac{1}{\rho_{i}^{2}}\big( 1+ \mathcal{W} ( s_{i} \rho_{i}^{2}) \big) +\dfrac{1}{2}\dfrac{1}{1+ \mathcal{W} ( s_{i} \rho_{i}^{2} ) }\Big) \frac{\partial}{\partial \lambda_{i}} \mathcal{W}( s_{i} \rho_{i}^{2}) \\
   &= y_{i} - \Big( \dfrac{1}{\rho_{i}^{2}}\big( 1+ \mathcal{W} ( s_{i} \rho_{i}^{2}) \big) +\dfrac{1}{2}\dfrac{1}{1+ \mathcal{W} ( s_{i} \rho_{i}^{2} ) }\Big) \frac{\mathcal{W} ( s_{i} \rho_{i}^{2} )}{1 + \mathcal{W} ( s_{i} \rho_{i}^{2} )} \frac{1}{s_{i} \rho_{i}^{2}}\frac{\partial}{\partial \lambda_{i}}( s_{i} \rho_{i}^{2}) \\
   &=  y_{i} - \mathcal{W} ( s_{i} \rho_{i}^{2} )  \Big( \dfrac{1}{\rho_{i}^{2}} + \frac{1}{2} \big( 1 + \mathcal{W} ( s_{i} \rho_{i}^{2} ) \big)^{-2} \Big) \; ,
\end{split}
\end{equation*}
and similarly
\begin{equation*}
\begin{split}
   \frac{\partial}{\partial \rho_{i}^{2}} \log \widetilde{Z}_{h_{i}} &= \frac{\partial}{\partial \rho_{i}^{2}} \Big( \frac{1}{2} \rho_{i}^{2} y_{i}^{2} -\dfrac{1}{\rho_{i}^{2}}\mathcal{W}( s_{i} \rho_{i}^{2}) -\dfrac{1}{2\rho_{i}^{2}}\mathcal{W}( s_{i} \rho_{i}^{2}) ^{2}-\dfrac{1}{2}\log \big( 1+ \mathcal{W} ( s_{i} \rho_{i}^{2}) \big) \Big) \\
   &= \frac{y_{i}^{2}}{2} + \frac{1}{\rho_{i}^{2}} \Big( \dfrac{1}{\rho_{i}^{2}}\mathcal{W}( s_{i} \rho_{i}^{2}) + \dfrac{1}{2\rho_{i}^{2}}\mathcal{W}( s_{i} \rho_{i}^{2}) ^{2} \Big) \\
   & \qquad - \Big( \dfrac{1}{\rho_{i}^{2}}\big( 1+ \mathcal{W} ( s_{i} \rho_{i}^{2}) \big) +\dfrac{1}{2}\dfrac{1}{1+ \mathcal{W} ( s_{i} \rho_{i}^{2} ) }\Big) \frac{\partial}{\partial \rho_{i}^{2}} \mathcal{W}( s_{i} \rho_{i}^{2}) \\
   &= \frac{y_{i}^{2}}{2} + \frac{1}{\rho_{i}^{2}} \dfrac{\mathcal{W}( s_{i} \rho_{i}^{2})}{\rho_{i}^{2}} \Big( 1 + \dfrac{1}{2}\mathcal{W}( s_{i} \rho_{i}^{2}) \Big) \\
   & \qquad - \Big( \dfrac{1}{\rho_{i}^{2}}\big( 1+ \mathcal{W} ( s_{i} \rho_{i}^{2}) \big) +\dfrac{1}{2}\dfrac{1}{1+ \mathcal{W} ( s_{i} \rho_{i}^{2} ) }\Big) \frac{\mathcal{W} ( s_{i} \rho_{i}^{2} )}{1 + \mathcal{W} ( s_{i} \rho_{i}^{2} )} \frac{1}{s_{i} \rho_{i}^{2}}\frac{\partial}{\partial \rho_{i}^{2}}( s_{i} \rho_{i}^{2}) \\
   &= \frac{y_{i}^{2}}{2} + \frac{1}{\rho_{i}^{2}} \dfrac{\mathcal{W}( s_{i} \rho_{i}^{2})}{\rho_{i}^{2}} \Big( 1 + \dfrac{1}{2}\mathcal{W}( s_{i} \rho_{i}^{2}) \Big) + \frac{1 + \rho_{i}^{2} y_{i}}{\rho_{i}^{2}} \Big( \frac{\partial}{\partial \lambda_{i}} \log \widetilde{Z}_{h_{i}} - y_{i} \Big)
\end{split}
\end{equation*}

\subsubsection*{Approximation of the log-normal Laplace transform from \citet{rossberg2008accurate}}

Being based on Laplace's method, the above approximation $\widetilde{\mathcal{L}}(s_{i})$ is expected to be accurate in the regimes with large $s_{i} = \exp{(\lambda_{i} + \rho_{i}^{2} y_{i})}$.
Coherently with this anticipation, our empirical analysis suggests that the scheme by \citet{asmussen2014laplace} ceases to be sufficiently accurate for observations with $y_{i}=0$, provided that $\lambda_{i}$ is not large enough.
Accordingly, in the present section, we introduce an alternative approximation scheme of the log-normal Laplace transform to deal with such cases.
\citet{rossberg2008accurate} rewrite such Laplace transforms in terms of a 2-fold convolution involving the Gaussian cdf, and then considers a $M^{th}$ order Taylor expansion of the latter.
Let $\mathcal{M}(s) = \E \big[ e^{s Z}\big] = \E \big[ e^{s \varphi e^{\rho X_0} }\big]$ be the moment generating function of a log-normal $Z=\varphi e^{\rho X_0}$, with $X_0$ being a standard normal.
After performing a change of variable, the authors rewrite the target function as
\begin{equation*}
\begin{split}
    \mathcal{G}(x) &\coloneq \mathcal{M}(-e^{-\rho x}/\varphi) = \int_{-\infty}^{\infty} \rho \, e^{\rho(x-x')} \, e^{-e^{\rho(x-x')}} \, \Phi(x') \, d x' \; .
\end{split}
\end{equation*}
Expanding $\Phi(x')$ around $x$ and exchanging the order sum and integral one has 
\begin{equation}\label{eq_F_rossberg}
\begin{split}
        \mathcal{G}(x) &\approx \widetilde{\mathcal{G}}(x)=\sum_{m=0}^M \frac{a_m}{(-\rho)^m} \Phi^{(m)}(x) + \mathcal{O}\left(\frac{1}{\rho^{M+1}} \right) \,
\end{split}
\end{equation}
where $\Phi^{(m)}(\cdot) $ is the $m^{th}$ derivative of the standard normal cdf, while $M$ is a positive integer.
The involved coefficients read $a_m=(-1)^m / m! \int_0^\infty d u \; e^{-u} (\ln u)^m$.
Such approximation is asymptotically exact in $\rho$, but not in $M$.
The authors prove a theoretical argument suggesting $M=6$ as a reasonable truncation value.
Following the authors, we rewrite equation~\eqref{eq_F_rossberg} in terms of the error function and its derivatives
\begin{equation*}
\begin{split}
    \operatorname{erf}(x)&= \frac{2}{\sqrt{\pi}} \int_{-\infty}^x e^{-w^{2}} \, d w  \qquad \qquad \operatorname{erfc}(x)=1-\operatorname{erf}(x) \\
    \operatorname{erf}^{(m)}(x)&= 2 \pi^{-1/2} (-1)^{m-1} H_{m-1}(x) e^{-x^{2}} \; ,
\end{split}
\end{equation*}
so that 
\begin{equation*}
\begin{split}
    \widetilde{\mathcal{G}}(x) &= \frac{1}{2} + \frac{1}{2} \operatorname{erf}\left( \frac{x}{\sqrt{2}} \right) - \frac{a_M}{2} \exp\left(\frac{\rho^{2}}{2} -x \rho\right) \operatorname{erfc}\left( \frac{\rho-x}{\sqrt{2}} \right) \\
    & \qquad + \exp\left(-\frac{x^{2}}{2}\right) \sum_{m=1}^{M-1}\frac{a_M-a_m}{\sqrt{\pi} \big(\rho \sqrt{2}\big)^{m}} H_{s-1}\left(\frac{y}{\sqrt{2}}\right) +\mathcal{O}\left(\frac{1}{\rho^{
    M}} \right) \; .
\end{split}
\end{equation*}
Here $ H_{m}(x)$ denotes the $m^{th}$ Hermite polynomials, which benefits from the two useful properties
\begin{equation*}
\begin{split}
H_{m+1}(x) = 2 \, x \, H_{m}(x) - 2 \, m \, H_{m-1}(x) \qquad\qquad 
H'_{m}(x) = 2 \, m \, H_{m-1}(x) \;.
\end{split} 
\end{equation*}
Accordingly, it is easy to show that
\begin{equation*}
\begin{split}
    \widetilde{\mathcal{G}}'(x) &= \exp\left(-\frac{x^{2}}{2}\right) \Bigg[ \, \frac{1}{\sqrt{2 \pi}}    +
    \frac{a_M}{2} \left( \exp\left(\frac{(\rho - x)^{2}}{2}\right) \rho \operatorname{erfc}\left(\frac{\rho - x}{\sqrt{2}}\right) - \frac{\sqrt{2}}{\sqrt{\pi}} \right) \\
    & \qquad - \frac{x}{\sqrt{\pi}} \frac{a_M-a_1}{\rho \sqrt{2}} - \frac{1}{\sqrt{2 \pi}} \sum_{m=2}^{M-1}\frac{a_M-a_m}{\big(\rho \sqrt{2}\big)^m} H_m\left( \frac{y}{\sqrt{2}}\right) \Bigg] 
     \\
  \widetilde{\mathcal{G}}''(x) &= \exp\left(-\frac{x^{2}}{2}\right) \Bigg[-\frac{x}{\sqrt{2 \pi}} - \frac{a_M}{2} \left( \exp\left(\frac{(\rho - x)^{2}}{2}\right) \rho^{2} \operatorname{erfc}\left(\frac{\rho - x}{\sqrt{2}}\right) - (\rho + x) \frac{\sqrt{2}}{\sqrt{\pi}} \right)  \\
    & \qquad - \frac{(1-x^{2})}{\sqrt{\pi}} \frac{a_M-a_1}{\rho \sqrt{2}} +
    \frac{1}{\sqrt{2 \pi}} \sum_{m=2}^{M-1}\frac{a_M-a_m}{\big(\rho \sqrt{2}\big)^m} H_{m+1}\left( \frac{y}{\sqrt{2}}\right) \Bigg] \; .
\end{split}
\end{equation*}

\subsubsection*{Adaptation to approximate moments matching in Poisson regression}

Similar to what was shown before, it is easy to see the hybrid normalizing constant in the Poisson case can be rephrased as 
\begin{equation*}
    Z_{h_{i}} = \frac{1}{y_{i} !} \, e^{y_{i} \lambda_{i} + \frac{1}{2} \rho_{i}^{2} y_{i}^{2}} \mathcal{G}\big(-\lambda_{i} /\rho_{i} - y_{i} \rho_{i} \big) .
\end{equation*}
Approximate moment matching can proceed once more exploiting the tractable approximation $\widetilde{\mathcal{G}}(\cdot)$ and its derivatives.
Notice that in the previous section, we worked out the details for the second derivative of $\widetilde{\mathcal{G}}(\cdot)$ as well, other than the first one.
This is because of an equivalence formulation of the exact version $\Delta_{i}$ as 
\begin{equation*}
     \Delta_{i} = 2 \frac{1}{Z_{h_{i}}} \frac{\partial Z_{h_{i}}}{\partial (\rho_{i}^{2})} - \Big( \frac{1}{Z_{h_{i}}} \frac{\partial Z_{h_{i}}}{\partial \lambda_{i}} \Big)^{2} = \frac{1}{ Z_{h_{i}}} \frac{\partial^{2} Z_{h_{i}}}{\partial \lambda_{i}^{2}} - \Big( \frac{1}{Z_{h_{i}}} \frac{\partial Z_{h_{i}}}{\partial \lambda_{i}} \Big)^{2} \; .
\end{equation*}
In the case of the approximation from \citet{rossberg2008accurate}, we found that approximating the second expression leads to more stable and accurate results, compared to targeting the first formulation.
The required approximate derivatives take the form
\begin{equation*}
\begin{split}
    \frac{\partial Z_{h_{i}}}{\partial \lambda_{i}} = Z_{h_{i}} \left( y_{i} - \frac{1}{\rho_{i}} \frac{\widetilde{\mathcal{G}}'\big(s
_{i} \big)}{\widetilde{\mathcal{G}}\big(s_{i} \big)}\right) \qquad \;
    \frac{\partial^{2} Z_{h_{i}}}{\partial \lambda_{i}^{2}} = Z_{h_{i}} \left( y_{i}^{2} - 2 y_{i} \frac{1}{\rho_{i}} \frac{\widetilde{\mathcal{G}}'\big(s
_{i} \big)}{\widetilde{\mathcal{G}}\big(s_{i} \big)} + \frac{1}{\rho_{i}^{2}} \frac{\widetilde{\mathcal{G}}''\big(s
_{i} \big)}{\widetilde{\mathcal{G}}\big(s_{i} \big)} \right)\; ,
\end{split}
\end{equation*}
where $s_{i}=-\lambda_{i} /\rho_{i} - y_{i} \rho_{i}$.

\section{Running times to compute the posterior moments only}\label{sec: time mom}

In this section, we report the running times to obtain the posterior moments only (i.e.\ excluding the time needed to compute the predictive probabilities) of the different methods presented in Sections \ref*{sec: simstudies} and \ref*{sec: illustration}.

\subsection*{Simulation study of Section \ref*{subsec: simProbit}}
\begin{table*}[ht]
\centering
\caption{\footnotesize{Running time, in seconds, to compute posterior moments with the \textsc{ep} approximation as in Algorithms \ref*{algo1} and \ref*{algo2} (\textsc{ep-eff}), with the \textsc{ep} approximation computed via the \texttt{R} function \texttt{EPprobit} (\texttt{EPprobit}), with the \textsc{pfm-vb} approximation (\textsc{pfm-vb}) and with \texttt{rstan} implementation of No-U-Turn \textsc{hmc} sampler (\textsc{nuts}) for probit regression with $n=500$ and $\nu^{2}=25$.
}}
\label{tab: simProbitOnlyMoments}
\begin{tabular}[c]{ll|ccccc}
 \multicolumn{2}{l|}{}  &  \multicolumn{5}{c}{\textit{p}}   \\
\hline
 & \textit{Method} & 125 & 250 & 500 & 1000 & 2000 \\ 
   \hline
Running time &\ \textsc{ep-eff} &\ 0.53 &\ 1.11 &\ 1.98 &\ 3.60 &\ 7.55 \\
(seconds) &\ \texttt{EPprobit}\ &\ 6.04 &\ 38.36 &\ 232.88 &\ 1867.20 &\ 21205.83 \\
&\ \textsc{pfm-vb}\ &\ 2.20 &\ 1.24 &\ 2.56 &\ 0.39 &\ 0.68 \\
&\ \textsc{nuts}\ &\ 151.87 &\ 490.64 &\ 1970.62 &\ 7743.65 &\ 20411.51 \\
\hline
\end{tabular}
\end{table*}

\newpage
\subsection*{Simulation study of Section \ref*{subsec: simPoisson}}
\begin{table*}[ht]
\centering
\caption{\footnotesize{Running time, in seconds, to compute posterior moments for $\tilde{n}=100$ test observations with the \textsc{ep} approximation as in Algorithms \ref*{algo1} and \ref*{algo2} (\textsc{ep-eff}), with the Metropolis-Hastings sampler from the \texttt{R} package \texttt{bpr} (\texttt{bpr}) and with the \texttt{rstan} implementation of the No-U-Turn \textsc{hmc} sampler (\textsc{nuts}) for Poisson regression with $n=500$ and $\nu^{2}=250/p$.
}}
\label{table: simPoissonOnlyMoments}
\begin{tabular}[c]{ll|ccccc}
 \multicolumn{2}{l|}{}  &  \multicolumn{5}{c}{\textit{p}}   \\
\hline
 & \textit{Method} & 125 & 250 & 500 & 1000 & 2000 \\ 
   \hline
Running time &\ \textsc{ep-eff} &\ 0.94 &\ 2.48 &\ 3.12 &\ 5.51 &\ 11.89 \\
(seconds) &\ \texttt{bpr}\ &\ 164.53 &\ 727.07 &\ 4021.05 &\ 25369.52 &\ 184152.96 \\
&\ \textsc{nuts}\ &\ 61.99 &\ 138.49 &\ 580.56 &\ 1082.21 &\ 1671.58 \\
\hline
\end{tabular}
\end{table*}

\FloatBarrier

\subsection*{Illustration of Section \ref*{subsec: illProbit}}
\begin{table*}[ht]
\centering
\caption{\footnotesize{Running time, in seconds, to compute posterior moments with the Monte Carlo sampler (\textsc{mc}), with the \texttt{rstan} implementation of the No-U-Turn \textsc{hmc} sampler (\textsc{nuts}), with the \textsc{ep} approximation as in Algorithms \ref*{algo1} and \ref*{algo2} (\textsc{ep-eff}), with the \textsc{ep} approximation computed via the \texttt{R} function \texttt{EPprobit} (\texttt{EPprobit}), and with the \textsc{pfm-vb} approximation (\textsc{pfm-vb}) for probit regression on the Alzheimer dataset ($n=300$) with $\nu^{2}=25$.
}}
\label{table: illProbitOnlyMoments}
\begin{tabular}[c]{l|ccccc}
 &  \multicolumn{3}{c}{Method}   \\
\hline
& \textsc{nuts} & \texttt{EPprobit} & \textsc{mc} & \textsc{ep-eff} & \textsc{pfm-vb}\\
\hline
 Running time (in seconds) & > 21600 & > 21600 & 1536.54 & 11.81 & 1.23\\ 
\hline
\end{tabular}
\end{table*}

\FloatBarrier

\subsection*{Illustration of Section \ref*{subsec: illPoisson}}
Running time HMC: 46153.91 secs.
Running time EP: 19.48 secs.

\printbibliography
\end{document}